\documentclass[11pt]{article}
\usepackage{comment}
\usepackage{amsmath,amssymb,graphicx}
\usepackage[version=3]{mhchem}
\usepackage{xcolor}

\definecolor{sbblue}{rgb}{0.0, 0.0, 1.0}

\usepackage[algoruled,boxed,lined]{algorithm2e}
\def\bx{{\mathbf x}}
\def\bX{{\mathbf X}}
\def\by{{\mathbf y}}

\usepackage{caption}

\usepackage[normalem]{ulem}

\usepackage[authoryear]{natbib}
\usepackage{subfigure}
\usepackage{url}

\begin{document}
\title{Linking stability with molecular geometries of perovskites and lanthanide richness using machine learning methods}                      



%
\author{Sampreeti Bhattacharya\thanks{Department of Chemistry, University of North Carolina at Chapel Hill, 
    Chapel Hill,
   NC, 
  USA}
   \\and\\Arkaprava Roy\thanks{Department of Biostatistics,
University of Florida,
    Gainesville,
  FL, 
    USA}}









\maketitle

{\bf Abstract:}
Oxide perovskite materials of type \ce{ABO3} have a wide range of technological applications, such as catalysts in solid oxide fuel cells and as light-absorbing materials in solar photovoltaics.
These materials often exhibit differential structural and electrostatic properties through their varied A-sites and B-sites especially if they are derived from lanthanides. Although, experimental and/or computational verification of these differences are often difficult and expensive. In this paper, we thus take a data-driven approach.
Specifically, we run three analysis using the dataset \cite{li2018data}  applying advanced machine learning tools to perform nonparametric regressions and also to produce data visualizations using latent factor analysis (LFA) and principal component analysis (PCA). We also implement a nonparametric feature screening step while performing our high dimensional regression analysis, ensuring robustness in our results.




\maketitle

\section{Introduction}

Perovskites are fascinating systems due to their intriguing physicochemical features and capacity to boost a wide range of electromagnetic and thermal processes. Owing to their ferromagnetic, ferroelectric, and piezoelectric characteristics, as well as ion conductivity, photocatalysis, and superconductivity applications, perovskite systems have a wide range of application \citep{hayward2002hydride,jin2008high,yamada2008perovskite,belik2016low}. These materials are potently versatile in optoelectronic and spintronic applications, and their formation and stability is governed by atomistic metrics such as tolerance factor ($t$) and octahedral factor ($\mu$), where 
$t=\left(r_{\mathrm{A}}+r_{\mathrm{X}}\right) / {\sqrt{2}\left(r_{\mathrm{B}}+r_{\mathrm{X}}\right)}$ and  $\mu=\left(r_{\mathrm{B}}/r_{\mathrm{X}}\right)$ respectively \citep{goldschmidt1926gesetze}.  For stable perovskite compounds, these $t$ values are in the range of 0.75 to 1.05, and $\mu$  values range from 0.00 to 0.15.
\citep{filip2018geometric,zhao2021combinatory}. In addition to traditional geometric descriptors, significant advancements in stable perovskite predictions are also achieved utilizing \cite{bartel2019new} data-driven 1D descriptor, like $\tau$  providing 92\% prediction performance for the ground-state-stable perovskites.

 In \ce{ABO3} type oxide perovskite systems, the 12-fold coordinated A-sites are occupied by large-size alkali metal ions, alkaline earth metal ions, or lanthanide cations, while the 6-fold coordinated B-sites are occupied by transition metals \citep{wexler2021factors}. 
The chemical diversity in oxide perovskites makes the class of compound extremely tunable \citep{gopalakrishnan2020perovskite, li2022chemical}, at the same time raising an issue of incorrect identification of formable systems and the associated stability \citep{li2017chemically}. In this work, we study a mixture of \ce{ABO3} perovskite systems \citep{bendersky2003transmission,liang2008synthesis,nag2014oxide}. Varied A and B sites cations in ternary, quaternary and sextenary types of compositions yield different geometric environments. This further affects the connected octahedra and thereby structural features like Shannon radii, distance between A/B and oxides and other such attributes \citep{zhou2020structural}. The multitude of possible combinations in these metal cations leads to varying structural dependence on thermodynamic and electronic properties.  

The studies in  \cite{giaquinta1994structural, goudochnikov2007correlations, filip2018geometric, liang2020electronic} demonstrate the efficacy of structural predictors in predicting the stability of perovskites. According to  \cite{filip2018geometric}, these metrics can achieve an accuracy as high as 80\% in predicting stability and formable perovskites. Albeit, these identifiers are susceptible to electrostatic interactions such as Jahn-Teller (JT) distortions \citep{hong2021local} and thermal motions of the crystal geometry. Alongside, spatial arrangement of \ce{BO6} octahedra and octahedral rotations are also linked with Glazer modes \citep{glazer1972structure,aleksandrov2001structural}. These geometric modes are capable of tuning the global and local symmetry of the crystal, thereby making structural distortions a crucial descriptor in identifying significant electronic properties \citep{jia2020persistent}. 

In the recent work by \cite{hong2021local}, lanthanide-based inorganic perovskites show structural attributes induced by local charge separation. 
The authors explained that the JT effect contributes to the spontaneous lengthening of bonds caused by the tetragonal crystal field. Similarly, octahedral tilts and rotations of oxygen octahedra also play a crucial role in influencing the stability, formability, magnetic, dielectric, and catalytic properties \citep{glazer1972structure, woodward1997octahedral}. 
On the other hand,  thermodynamic stability of perovksite systems can generally be connected to stability and formability \citep{armiento2014high,jacobs2018material} through energy above the convex hull (EHull)  measured using the convex hull analysis \citep{liu2015spinel} and energy of formation (EForm) calculated using electronic structure theory such as density functional theory (DFT). As also shown in  \citep{li2018predicting}, EForm can be reliably linked to EHull when using a threshold value of 40 meV/atom.  However,
 limitations to theoretical prediction using DFT in identifying non-degenerate ground states for lanthanide materials make accurate electronic property calculations a challenging task \citep{ferbinteanu2017density,liu2020screening,shetty2022predicting}. As a result, data-driven approaches are gaining importance to perform a computational investigation of predicting stable perovskite systems \citep{sun2020accelerating}.
 
Data-driven approaches can also be utilized in understanding relevant features/descriptors arising due to specific atomic environments in perovskites systems.
Specifically, we are interested in the differential characteristics among these features, important for predicting stability across the lanthanide and non-lanthanide derived perovskites.

In this paper, we run three data-driven analyses. First, we investigate the relationship between structural features and physicochemical properties of species occupying A-sites and B-sites. 
We employ a nonparametric marginal screening algorithm \citep{xue2017robust} to first identify a subset of the most important descriptors for our analysis, and subsequently, we fit multivariate predictive models using the screened dataset.
Second, we study the effects of lanthanide atoms on the stability of perovskites after segregating the descriptors into two groups. The first group consists of the systems with at least one lanthanide atom in their atomic sites, characterizing them as lanthanum-containing substructures, and equivalently, non-lanthanum-containing substructures are defined as the systems without any lanthanide atoms. For this analysis, we only consider the subset of predictors that are identified in the screening process described in the first analysis. The rest of the article is organized as follows. 

In the next section, we describe the dataset.  Section~\ref{sec: Analysis plan} presents the description of models that are used in this paper. In Section~\ref{opdataanaly}, we describe our marginal feature screening algorithm followed by model fitting procedure. In Section~\ref{sec:resultanalysis} we present the results from our analysis, followed by some concluding remarks in Section~\ref{sec:conclusion}.
 

\vspace{-3mm}

\section{Oxide perovskite dataset} 
\label{sec: opdataanalys}
\begin{figure}[h]
    \centering
    \includegraphics[height=70 mm, width=160mm]{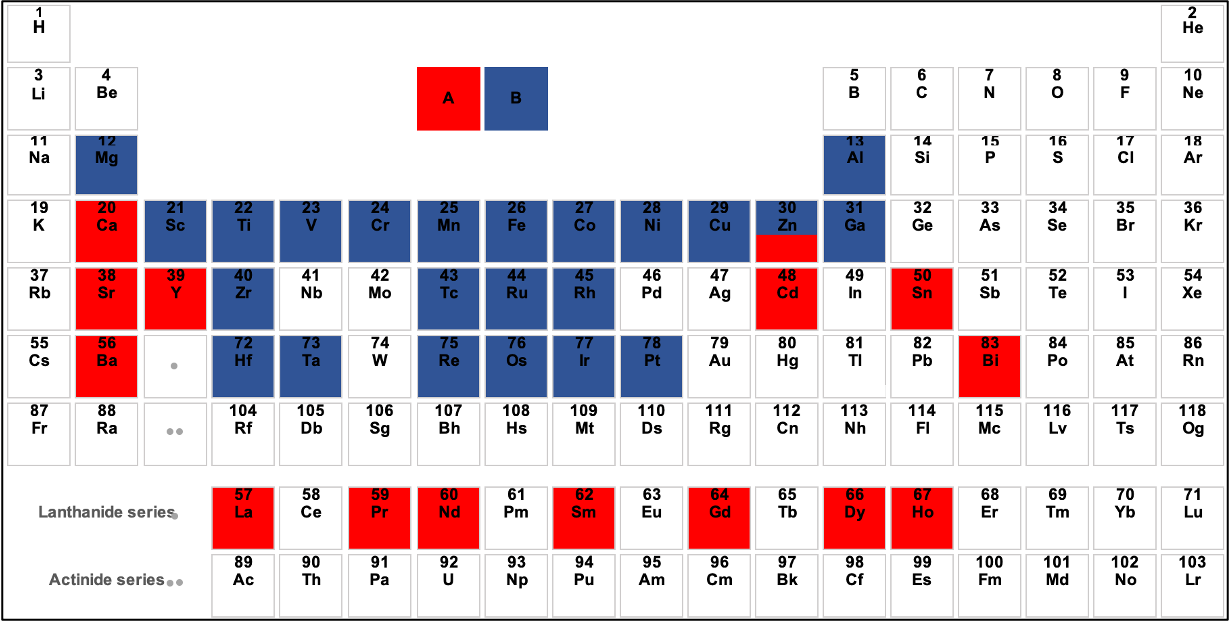}
    \caption{Distribution of A-site (shown in color red) and B-site (shown in color blue) cations across the periodic table. }
    \label{fig:perd_table}
\end{figure}

The dataset for the analysis is sourced from  \cite{li2018data}. 
It contains the elemental properties of 1929 oxide perovskites and their phase stability energies, generated from DFT and convex hull analysis. The distribution of A-site and B-site cations across the periodic table is shown in Figure \ref{fig:perd_table}. All the associated property measurements in the curated dataset are performed at a temperature of 1073 K and an oxygen partial pressure of 0.2 atm \citep{jacobs2018material}.
These settings mimic the approximate working conditions of solid oxide fuel cell (SOFC) cathodes and hence are a more realistic dataset to train and test the dependence of elemental properties. The dataset is composed of perovskite systems  comprising elements from a candidate set of \ce{A} cations: {Ba, La, Y, Pr, Gd, Dy, Ho, Nd, Sm, Ca, Sr, Bi, Cd, Sn, Zn}, \ce{B} cations such as {Fe, V, Cr, Mn, Sc, Co, Ti, Mg, Ni, Zr, Ga, Hf, Nb, Ta, Re, Tc, Ir, Os, Ru, Rh, Al, Cu, Pt, Zn}, and \ce{X}$^{2-}$ as O. In total, the dataset consists of 71 ternary (e.g. \ce{ABO3}), 1248 quaternary (e.g. \ce{A_nA$^\prime$_{1-n}BO3}), 601 quinary (e.g. \ce{A_nA$^\prime$_{1-n}B_mB$^\prime$_{1-m}O3}) and 9 sextenary perovskite oxide systems. 
The dataset comprises the descriptors that are created using an extensive elemental property database of physical and chemical properties of elements in their atomic form compiled from the Materials Agnostic Platform for Informatics and Exploration (MAGPIE) \citep{ward2016general} and the web chemical elements' database in Resources for Teaching Science to construct the matrix of features used to train our machine learning models. Further details about the data are available in  \cite{jacobs2018material, li2018data}. 

\vspace{-3mm}
\section{Analysis plan} 
\label{sec: Analysis plan}
Regression models provide a straightforward way to interpret relationships between one or more independent variables and an outcome variable. Here, we consider several regression models such as kernel ridge regression \citep{geron2017hands}, artificial neural networks \citep{mcculloch1943logical}, support vector regression (SVR) \citep{vapnik1999nature}, and random forest \citep{breiman2001random}, for our proposed analyses.
We discuss the formulation of the candidate regression models in the supplementary materials.
Although they are all explained for univariate response for simplicity, it requires minor modifications to the respective objective functions in case of multivariate response, which is accommodated using  multi-output-regressor. 
For clarity, we first define some of the notations.
We denote the response variable of $i$-$th$ sample as $y_i$, corresponding to the $p$-dimensional predictor vector $x_i$ for $i=1,\ldots,N$. In matrix notation, we denote $X$ as the design matrix, such that $i$-$th$ row of $X$ is $x_i$ and $Y$ as the response vector, such that $j$-$th$ element of $Y$ is $y_j$ . Thus, $X$ is $N\times p$ dimensional matrix and that $Y$ is an $N$-length vector.

For high dimensional regression, filtering out irrelevant descriptors is imperative before deploying any machine learning algorithm. Thus, from the available features, we screen a subset of important features.
We apply a nonparametric marginal screening method which relies on marginal associations between the response and each predictor separately.
The details are discussed in the later part of this section.
\subsection{Marginal feature screening}
\label{sec: Marginal feature}
 We apply a nonparametric screening method, proposed in  \cite{xue2017robust} as a pre-processing step before running our analyses with high dimensional structural features as predictors. It relies on the property that a zero correlation between the response and a given feature would mean their independence when the (response, feature) pair follows a bivariate normal distribution.
Hence, the joint covariance matrix of the pair will be diagonal if they are independent and jointly normal.
Here, joint normality is important, as marginal normality does not necessarily lead to joint normality.
The screening step thus tests bivariate normality with diagonal covariance between the response and each feature. 
The overall screening step relies on the following probabilistic argument. {If conditional distribution of $y$ given the predictors $x=(x_1,...,x_p)$ i.e. $P(y\mid x)$ functionally depends on $x_k$, then $y$ and $x_k$ are usually marginally dependent as well.
Hence, a relatively moderate-size set of variables can be constructed by selecting only the predictors
that are marginally dependent with $y$, and this procedure is commonly known as
independence screening. Furthermore, such a screening procedure is referred to as a {\it sure independence screening method} when it can be shown to identify all the important predictors in a supervised learning setting, with probability tending to 1 as sample size increases \citep{fan2008sure}. The method in \cite{xue2017robust} is shown to satisfy such property.}

The protocol mentioned in  \cite{xue2017robust} recommends first transforming the response and each predictor into normally distributed random variables by applying non-paranormal transformations \citep{liu2009nonparanormal}.
Subsequently, the Henze-Zirkler (HZ) test \citep{henze1990class} statistic is computed to statistically quantify the separation or dissimilarity between the bivariate empirical distribution and a bivariate normal distribution with diagonal covariance for each pair of (transformed response, one transformed predictor) separately.
A larger value of the HZ test statistic would thus indicate a stronger association between the response and the given predictor, as then their joint distribution does not follow a bivariate normal with diagonal covariance.
Hence, the predictors with large HZ test statistic values are included in the model. The necessary theoretical justifications of this procedure are provided in \cite{xue2017robust}.
In this paper, we propose a minor modification to the original algorithm for the purpose of handling multivariate responses. Our modified set of steps are in Algorithm~\ref{algo} and Figure \ref{fig:algo1}.

\subsection{Computational details}
\label{sec:computation}
In this work, we leverage a hybrid approach, incorporating widely used tools from R version 4.2.1 and Python 3.8.13. For the implementation of marginal feature screening, the {\tt{huge}} \citep{zhao2012huge} package in R is employed to obtain the non-paranormal transformations and to calculate the HZ-statistics \citep{henze1990class}. Multivariate regression analyses are executed using the corresponding regression classes provided by the {\tt{scikit-learn}} \citep{scikit-learn} package. To extend these models to accommodate multi-response multivariate regression, the {\tt{MultiOutputRegressor}} class from the {\tt{scikit-learn}} package is utilized. A comprehensive list of the associated parameters for each case is provided in Table \ref{table:1}. Principal Component Analysis (PCA) is conducted using the {\tt{prcomp}} package in R to compute the principal components, while linear factor analysis is performed using the {\tt{sklearn.decomposition.FactorAnalysis}} class, with {\tt{n\_component=10}} (the number of latent factors) specified, within the {\tt{scikit-learn}} package. We calculate the permutation importance from {\tt{sklearn.inspection}} setting {\tt{n\_repeats=20}} (the number of permutations) and {\tt{scoring=neg\_mean\_squared\_error}} (the error determining metric).

\section{Oxide perovskite dataset analysis}\label{opdataanaly}
\begin{figure}[h]
\centering
\includegraphics[height=70 mm, width=160mm]{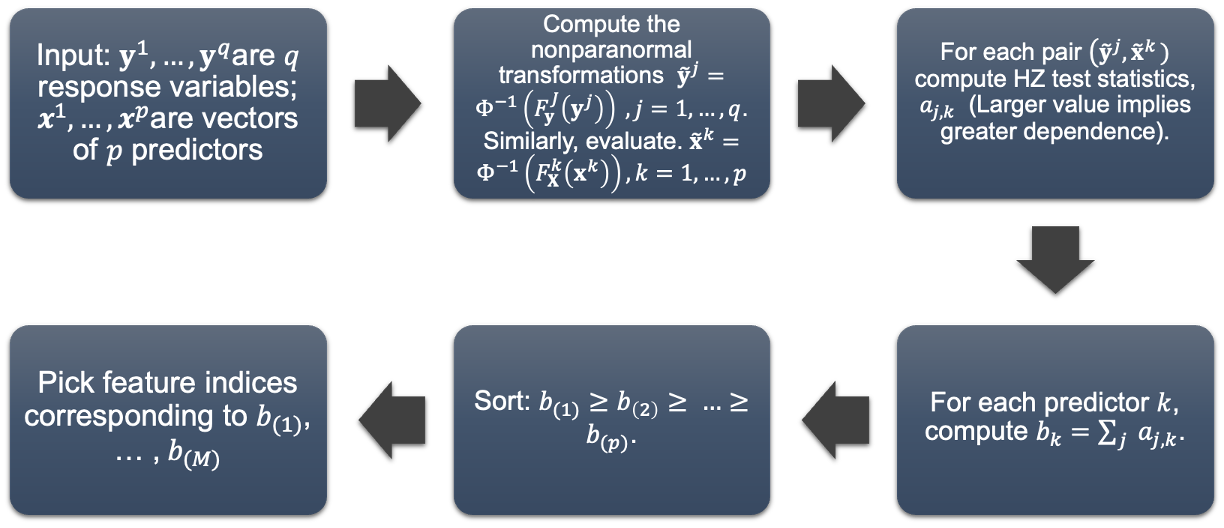}
       \caption{The workflow of the marginal screening method to select $M$ number of features out of $p $ predictors based on $q$ responses, as described in Algorithm \ref{algo}. In step 2, $\Phi(\cdot)=$ Cumulative distribution function of univariate normal and $F_{\by}^{j}(\cdot)=$ Cumulative distribution function of $\by^{j}$. }
\label{fig:algo1}
    \end{figure}

Before proceeding with our analysis, we apply min-max normalization for each feature vector. 
For high dimensional regressions, we apply the screening method to select the top 100 features, that are strongly related to the response. 
These 100 features are then used for different regression analyses. The steps are illustrated in Algorithm~\ref{algo} and Figure \ref{fig:algo1} to screen $M$-many most relevant predictors for $q$-dimensional response and $p$-dimensional predictor.

\begin{table}[h] 
\caption{The optimized parameter and hyperparameter list for each model used in univariate and multivariate multiple regressions and are optimized to get the best result.}
\label{table:1}
\centering
\resizebox{1.0\textwidth}{!}{%
\begin{tabular}{|c|c|c|}
\hline
Models                        & \begin{tabular}[c]{@{}c@{}}Parameters/Hyperparameters \\  for cross validation\end{tabular} & \begin{tabular}[c]{@{}c@{}}Constant \\ Parameters/Hyperparameters\end{tabular} \\ \hline\hline
Support Vector Regression & {\tt{gamma}}, {\tt{C}}                          & {\tt{kernel=rbf }}                                                  \\ \hline
Kernel Ridge Regression       & {\tt{alpha}}, {\tt{kernel}}, {\tt{gamma}} & -                                                                         \\ \hline
Neural Network                & {\tt{hidden\_layer\_sizes}},                                       & {\tt{activation=logistic}}, {\tt{solver=lbfgs}},   {\tt{alpha=0.001}},{\tt{learning\_rate=constant}} \\ \hline
Random Forest                 & {\tt{n\_estimators}},   {\tt{max\_samples}}     & {\tt{criterion=squared\_error}}                                                    \\ \hline
\end{tabular}}
\label{tab:methods}
\end{table}

\begin{algorithm}[htbp]
\SetAlgoLined
(i) Input: $\by^{1},\ldots,\by^{q}$ are $q$ response variables; $\bx^{1},\ldots,\bx^{p}$ are vectors of $p$ predictors;\\
(ii) Compute the nonparanormal transformations \citep{liu2009nonparanormal} $\tilde{\by}^{j}=\Phi^{-1}(F_{\by}^{j}(\by^{j}))$ for $j=1,\ldots, q$. Similarly, evaluate $\tilde{\bx}^{k}=\Phi^{-1}(F^{k}_{\bX}(\bx^{k}))$ for $k=1,\ldots,p$.

($\Phi(\cdot)=$ Cumulative distribution function of univariate normal and $F_{\by}^{j}(\cdot)=$ Cumulative distribution function of $\by^{j}$)\\

(iii) For each pair $(\tilde{\by}^{j}, \tilde{\bx}^{k})$, compute HZ test statistic $a_{j,k}$ which is used to test the hypothesis $H_0:(\tilde{\by}^{j}, \tilde{\bx}^{k})\sim\textrm{Normal}(0, I_2)$ (Larger value implies greater dependence)\\

(iv) For each predictor, get $b_k=\sum_j a_{j,k}$ \\

(v) Pick the top $M$ predictors  with the highest $b_k$-values.
  \caption{Marginal screening workflow}
 \label{algo}
\end{algorithm}

In our perovskite data application, we have $q=7$, $p=955$, and set $M=100$.
We can summarize our analyses primarily in three parts. First, we apply a linear factor model to the screened features obtained by applying the screening steps.
The factor model helps to uncover the underlying dependencies among them. Here, we are primarily interested in identifying the similarities and dissimilarities in their dependency structures across the lanthanum-containing and non-lanthanum-containing perovskites. 
Secondly, we identify the screened features, that are most relevant in explaining the seven geometric blueprints, which are  octahedral factor (`octahedral\_factor'),  ionic octahedral factor (`octahderal\_factor\_ionic'),  Goldschmidt tolerance factor(`goldschmidt\_TF\_factor'), 
 ionic Goldschmidt tolerance factor (`goldschmidt\_TF\_factor\_ionic'),  distance between cations A and oxygen atoms (`A\_O'),  distance between B and oxygen atoms (`B\_O'), and  the average distance between cations A and B (`A\_B'). Finally, we investigate the varying relationship of these seven geometric features with thermodynamic stability, as measured by EHull and EForm \citep{li2018predicting} across the lanthanum-containing and non-lanthanum-containing substructures.

As mentioned in the previous section, we use multivariate multiple regression for our first analysis. To identify the best model, we run a predictive study using the regression models described in Section~\ref{sec: Analysis plan}. We also fit linear regression models with penalties such as lasso, elastic net, and ridge. 
However, those results are omitted owing to their extremely poor performance in comparison to the other non-linear methods.

To compare across all the non-linear regression models mentioned in Table \ref{table:1} we consider the out-of-sample mean squared error (MSE). Out-of-sample MSE is one of the most popular metrics for model comparison, as it provides a model-independent assessment of the fit while facilitating efficient cross-method comparison.
To compute the MSE, we consider 20 random test-train splits of the data.
For each split, the training set consists of 80\% of the data and the remaining 20 \% is used to form the test set.
We train each model described in Section \ref{sec: Analysis plan} for every test-train split. In each split, the training set is used to estimate the model parameters for each model, enlisted in the second column of Table \ref{tab:methods} applying a 5-fold cross-validation technique \citep{refaeilzadeh2009cross}. Subsequently, we predict the responses utilizing the test data set for each model
and compute the out-of-sample MSE. MSEs are further averaged over all the 20 splits to obtain a single average, described as a pooled MSE estimate. The lowest pooled MSE is used to identify the candidate model most suitable for the data. The selected model is then reapplied to the complete data to quantify the importance of each feature. Specifically, we apply the permutation importance algorithm. The workflow of the entire analysis is elucidated below in Figure \ref{Fig:flowchart}

For our second analysis, we study the effect of the lanthanum-containing cationic sites on the stability and formability of perovskites by constructing a binary group vector setting `1' for perovskite with at least one lanthanide atomic site and `0' otherwise. The binary group vector helps in identifying the differential characteristics across lanthanum and non-lanthanum element-based substructures of data. It is partially motivated due to the following preliminary analysis.

\begin{figure*}[htbp]

    \centering
      \includegraphics[height=80 mm, width=115mm]{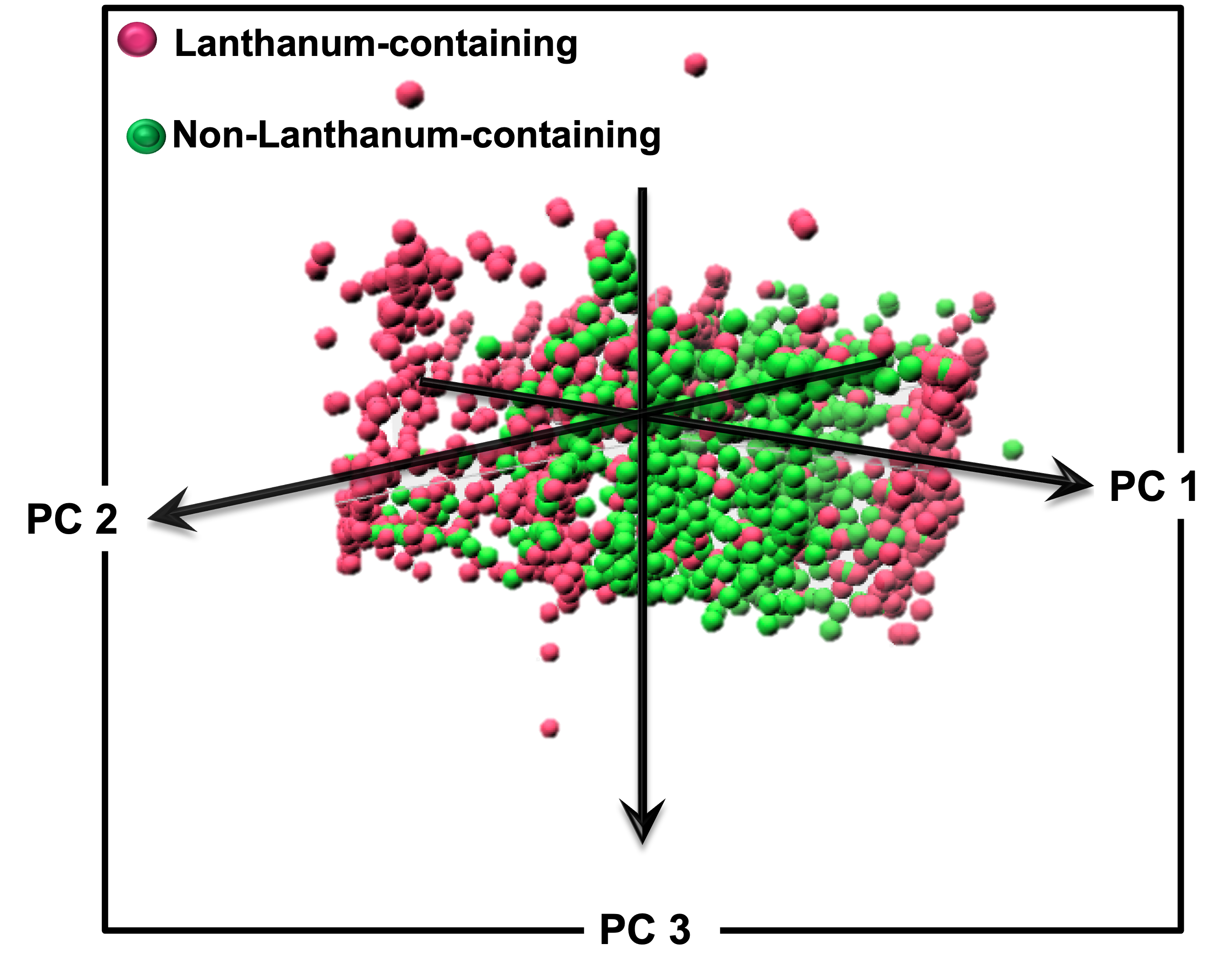}
      \caption{Principal components (PCs) are shown along PC1 and PC2. The pink color indicates non-lanthanum-containing substructure and the green color exhibits lanthanum-containing. Principal components are calculated using combined data, appending EHull and EForm profiles along with the seven geometric descriptors.}
      \label{Fig:pca1}
\end{figure*}

As an exploratory analysis, we run the principal component analysis with the combined set of seven geometric features along with EHull, EForm energy profiles and examine their variabilities across the two data substructures.
We show the principal components in Figures \ref{Fig:pca1} where the lanthanum-containing substructure illustrated using the pink dots are more clustered in comparison to the non-lanthanum derived features, illustrated using the green dots, which on the other hand are seen to be more dispersed. These principal components explain 99 \% variation of the data. In the following section, we conduct analyses to investigate additional distinctive characteristics of the substructures based on lanthanides.
\begin{figure}[h]
\centering
\includegraphics[height=100 mm, width=160mm]{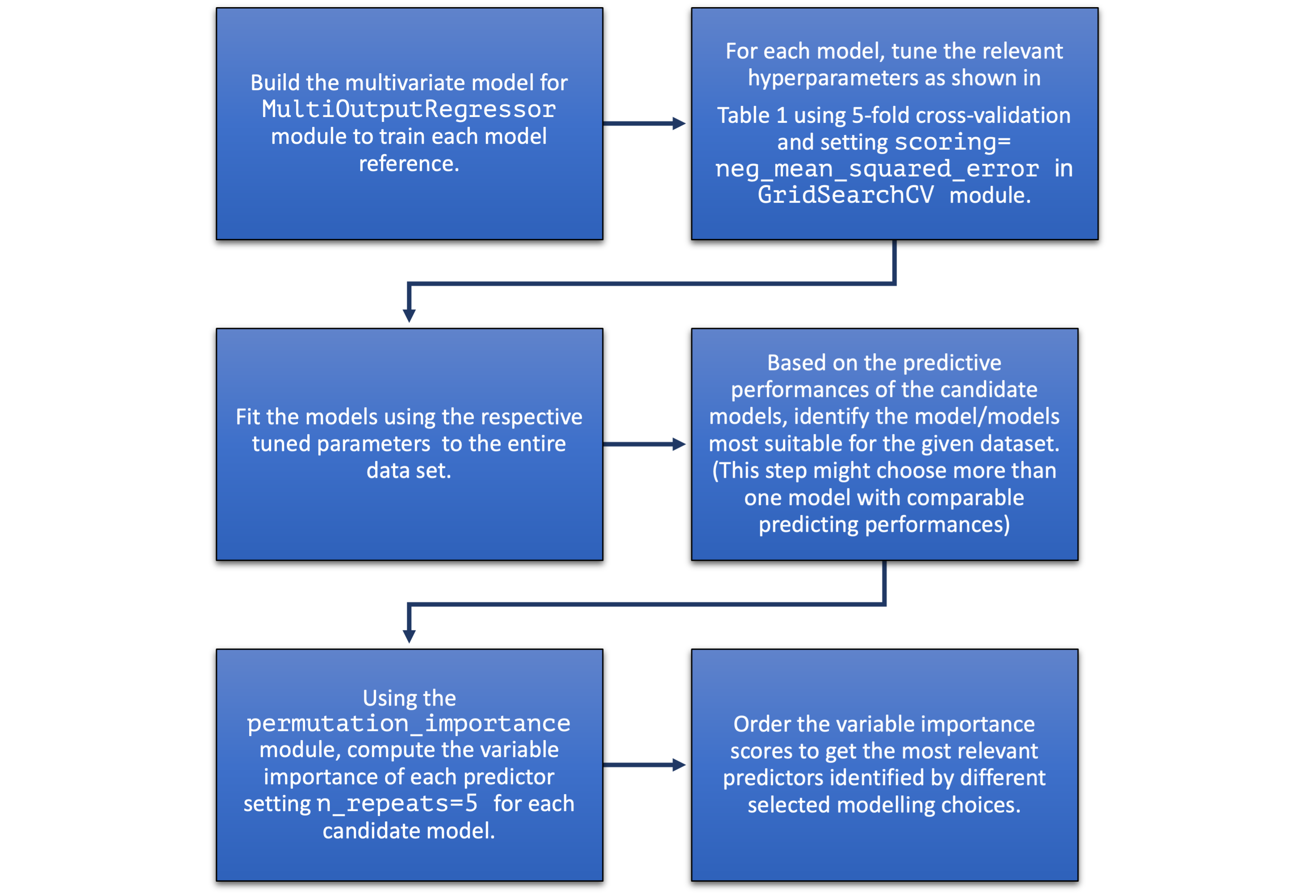}
       \caption{The workflow of the data analysis performed in this work. Based on the dimension of the response vector, the first box is used. For example, in the case of 1D responses, the {\tt{MultiOutputRegressor}} is not used. In the second box {\tt{GridSearchCV}} is performed for different sets of parameters that are used based on the model type as mentioned in Table \ref{table:1}. The rest of the steps are common irrespective of the model type, for {\tt{permutation\_importance}} the default scoring is used.}
\label{Fig:flowchart}
    \end{figure}

\section{Results and discussion} \label{sec:resultanalysis}
Our overall analysis has three key parts, which are discussed in the following three subsections in detail.

\subsection{Differential association of screened features on the lanthanide atomic site}
\begin{figure}[h]
    \centering
      \subfigure[]{\label{Fig:cor_l}\includegraphics[height=40 mm, width=62mm]{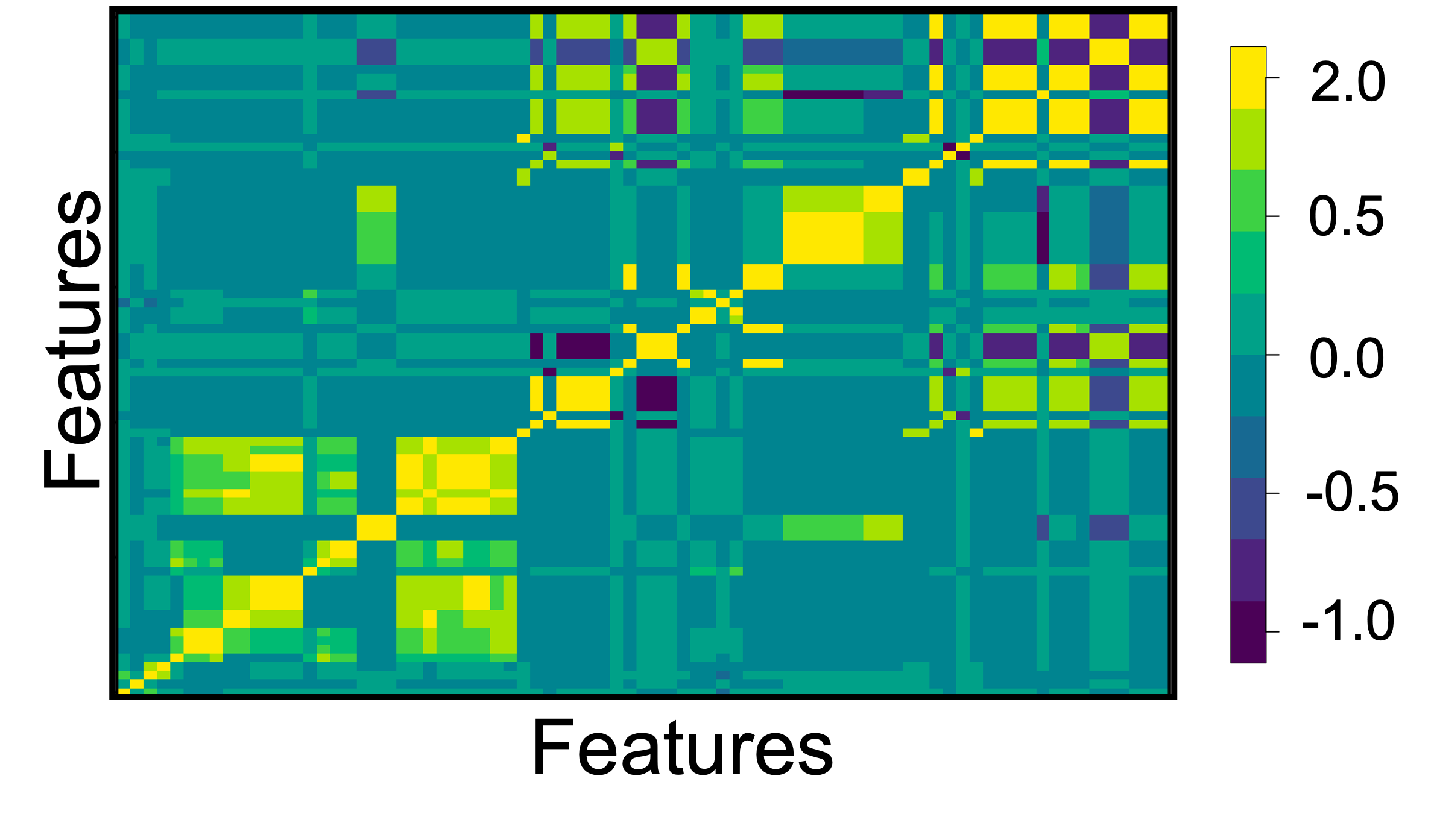}}
      \subfigure[]{\label{Fig:cor_nl}\includegraphics[height=40 mm, width=60mm]{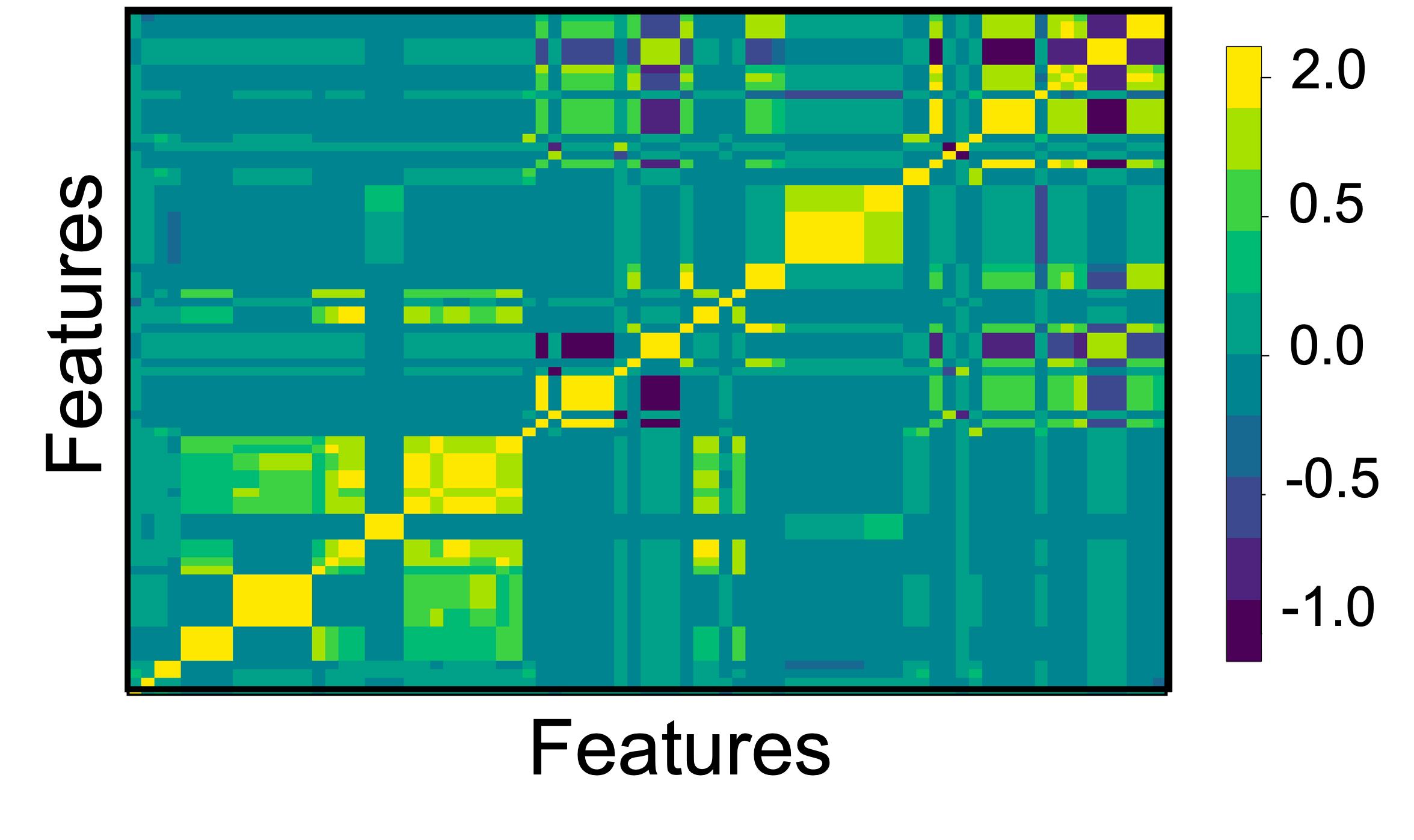}}
      \subfigure[Lanthanide substructure]{\label{Fig:fa_l}\includegraphics[height=40 mm, width=60mm]{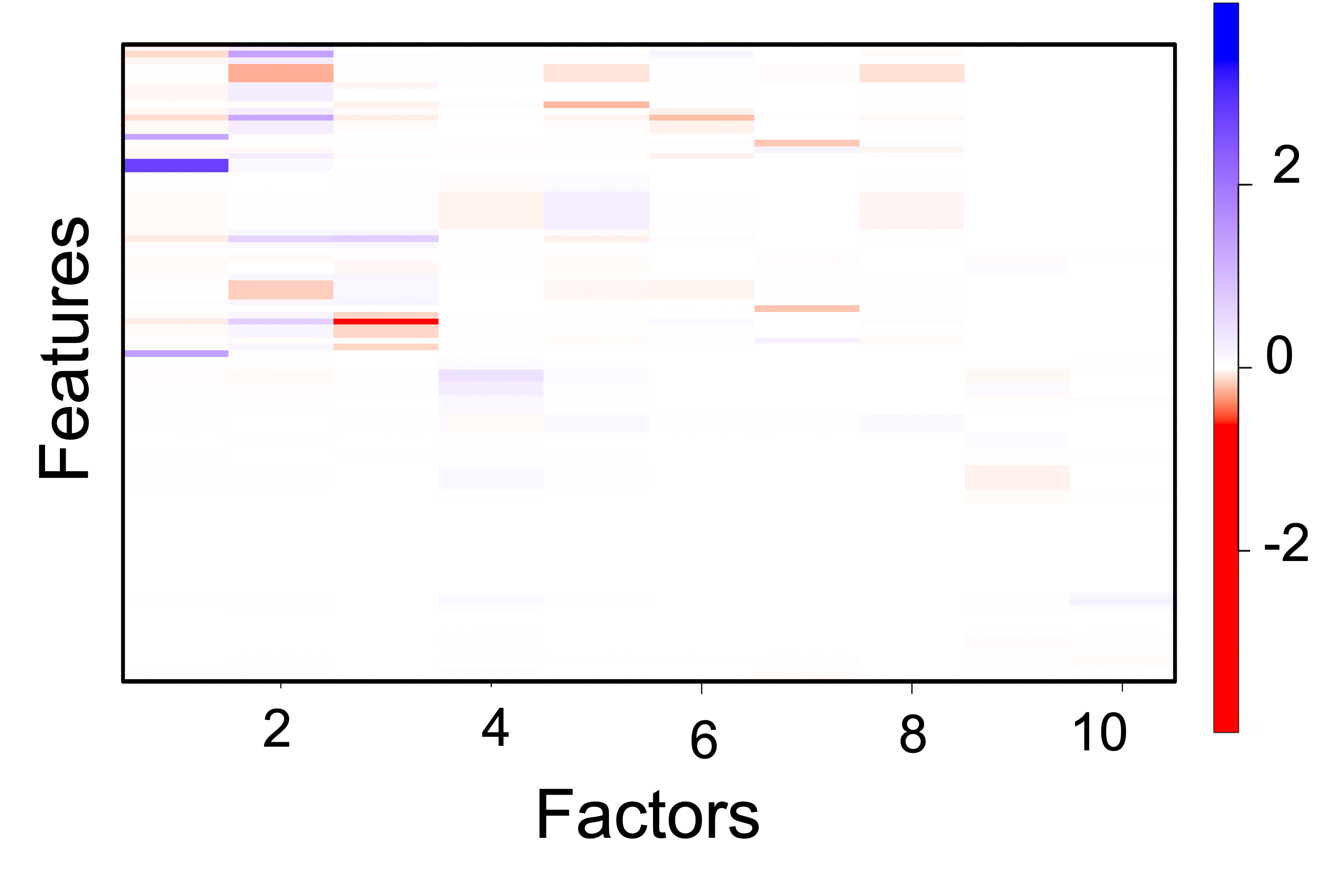}}
       \subfigure[Non-lanthanide substructure]{\label{Fig:fa_nl}\includegraphics[height=40 mm, width=60mm]{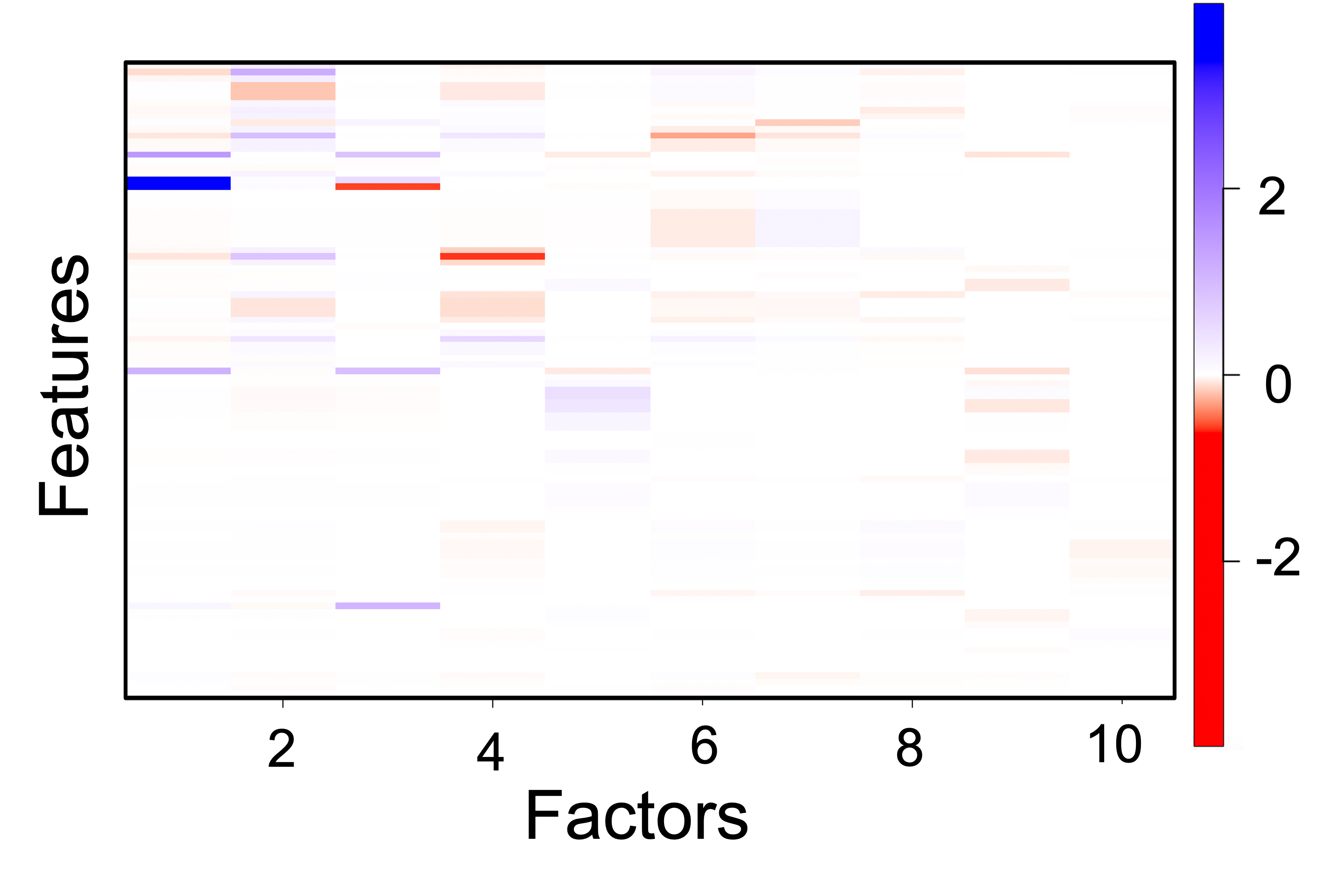}}
     \caption{Correlation matrices for the 100 screened features for (a) lanthanum-containing and (b) non-lanthanum-containing substructures. Corresponding linear factors matrices for (c) lanthanum-containing and (d) non-lanthanum-containing data fragments show the 10 estimated factors in the horizontal axes. For clarity the names of the variables are not reported in the figures and are thus clearly shown in Figures S1 and S2.}
    \label{fig:fa}
    \end{figure}
Geometric blueprints serve as significant features in explaining the stability and formability of perovskites \citep{filip2018geometric,bartel2019new}.
Our screening procedure helps to identify the top 100 elemental descriptors that are strongly associated with the blueprints. Since the stability profile varies through doping or substitutions at A- sites or B-sites  \citep{li2015high,wang2016scandium,schader2017piezoelectric,song2018compositional}, we analyze the differential associativity among the selected descriptors across the two substructures. 

For this analysis, we resort to linear factor modeling, which is a natural choice for relating the multivariate data to lower-dimensional characteristics while reducing the number of parameters needed to represent the covariance. The general formula for factor analysis used  is $ X=F\times\lambda+\epsilon$, where $X$ is the matrix of screened descriptors, $F$ is a matrix of factors,
$\lambda$ is a matrix of factor loading
and $\epsilon$ is a matrix of residuals. This method aims at getting a low-dimensional dependency pattern of the covariance \citep{bunte2016sparse}. 
Figure \ref{fig:fa} exhibits the correlation and the estimated factors of the screened features for the lanthanum-containing and non-lanthanum-containing data substructures, respectively. Although, the correlation matrices in Figures \ref{Fig:cor_l} and \ref{Fig:cor_nl} are relatively similar.
However, there are differences, identified in the factor loading ($\lambda$) matrices as in Figures \ref{Fig:fa_l} and \ref{Fig:fa_nl}, indicating differential dependency patterns post the second factor. The differences in the descriptors associated with the first five significant factors describing lanthanum-containing and non-lanthanum-containing data fragments are further shown in the insets of Figures S1 and S2 in the supplementary materials. 
The following subsection presents our regression analysis using these screened features. 

\subsection{Model assessment and relevant feature analysis}\label{subsec:5.2}
We now analyze the relationship between the seven geometric blueprints and the selected set of screened variables by applying different machine-learning methods. The best model will subsequently be identified following the protocol outlined in Figure~\ref{Fig:flowchart}. As shown in Table \ref{Table:2} the pooled mean for out-of-sample MSEs are the lowest for random forest and kernel ridge. Numerically, it is the random forest. However, the performance of the kernel ridge is almost equivalent. To obtain the most relevant features, we calculate the permutation importance setting with 50 repetitions. 
We find that weighted average A-site Shannon radii (`Asite\_shannon\_radii\_weighted\_avg'),       weighted average B-site Shannon radii     (`Bsite\_shannon\_radii\_weighted\_avg'),     weighted average B-site ionic radii      (`Bsite\_Ionic.Radius\_angstroms\_weighted\_avg') and average A-B site Shannon radii  (
`shannon\_radii\_AB\_avg') are the four most relevant features, selected by both random forest and kernel ridge unanimously, which also suggests robustness in our inference plan of Figure~\ref{Fig:flowchart}. 
The conclusion corroborates with the work by  \cite{johnsson2005crystallography}, where geometric features are governed primarily by the coordination environment and site symmetry of the atomic positions. 

 Previously reported studies in  \cite{attfield2002cation,richter2009materials} highlight that while many attributes of the perovskite systems result from the B cations, it is the  A-site cations that are greatly responsible for tuning chemical and/or electronic properties. The
radius of the A-site cation is essential  in controlling perovskite
properties, and in fact `Asite\_shannon\_radii\_weighted\_avg' is the most significant feature selected in this study. Similarly, the selection of `Bsite\_Ionic.Radius\_angstroms\_weighted\_avg' and   `Bsite\_shannon\_radii\_weighted\_avg' features the B-site ionic radius and is consistent with the results reported in \citep{hilpert2003defect}. Such dependency emphasizes that the expansion observed with the absence or presence of lanthanide possessing systems is mainly determined by the relative change of the average B-site ionic radius. Although, with a decrease of the average
radius of the A site ion, the \ce{BO6} octahedra starts to tilt along a particular Glazer mode, and reduces the excess space around the A-site cation
 \citep{cherif2002effect}.
Constriction of space results in a B-O-B bond angles less than 180 $^\circ$ while affecting the stability of the lattice. Owing to larger deviations in the A-site and B-site Shannon radii respectively, the effect of `shannon\_radii\_AB\_avg' is also one of the significant descriptors affecting the geometric blueprints.

\begin{table}[h] 
\centering
 \caption{Average out-of-sample MSEs over 20 random test-train splits of the data for the candidate regression models associated after regressing multivariate geometric descriptors on the 100 screened elemental and non-elemental descriptors. (the best performers are in bold).}
  \label{Table:2}
\resizebox{0.3\textheight}{!}{\begin{tabular}{|c | c| }
\hline
Methods        & \begin{tabular}[c]{@{}c@{}}Pooled Mean \\ MSE\end{tabular}  \\ \hline \hline
SVM            & 0.00279                                                                                                                   \\ \hline
Kernel Ridge   & {\bf 0.00046}                                                                                                                 \\ \hline
Neural Network &  0.00066                                                                                                                  \\ \hline
Random Forest  & {\bf 0.00044}                                                                                                                   \\ \hline
\end{tabular}}
\end{table}

 \subsection{Impact of geometric blueprints on EHull and EForm}
\begin{figure}[htbp]
    \centering
    \subfigure[p-value = $5.36\times10^{-23}$]{\label{Fig:gold}\includegraphics[height=30 mm, width=45mm]{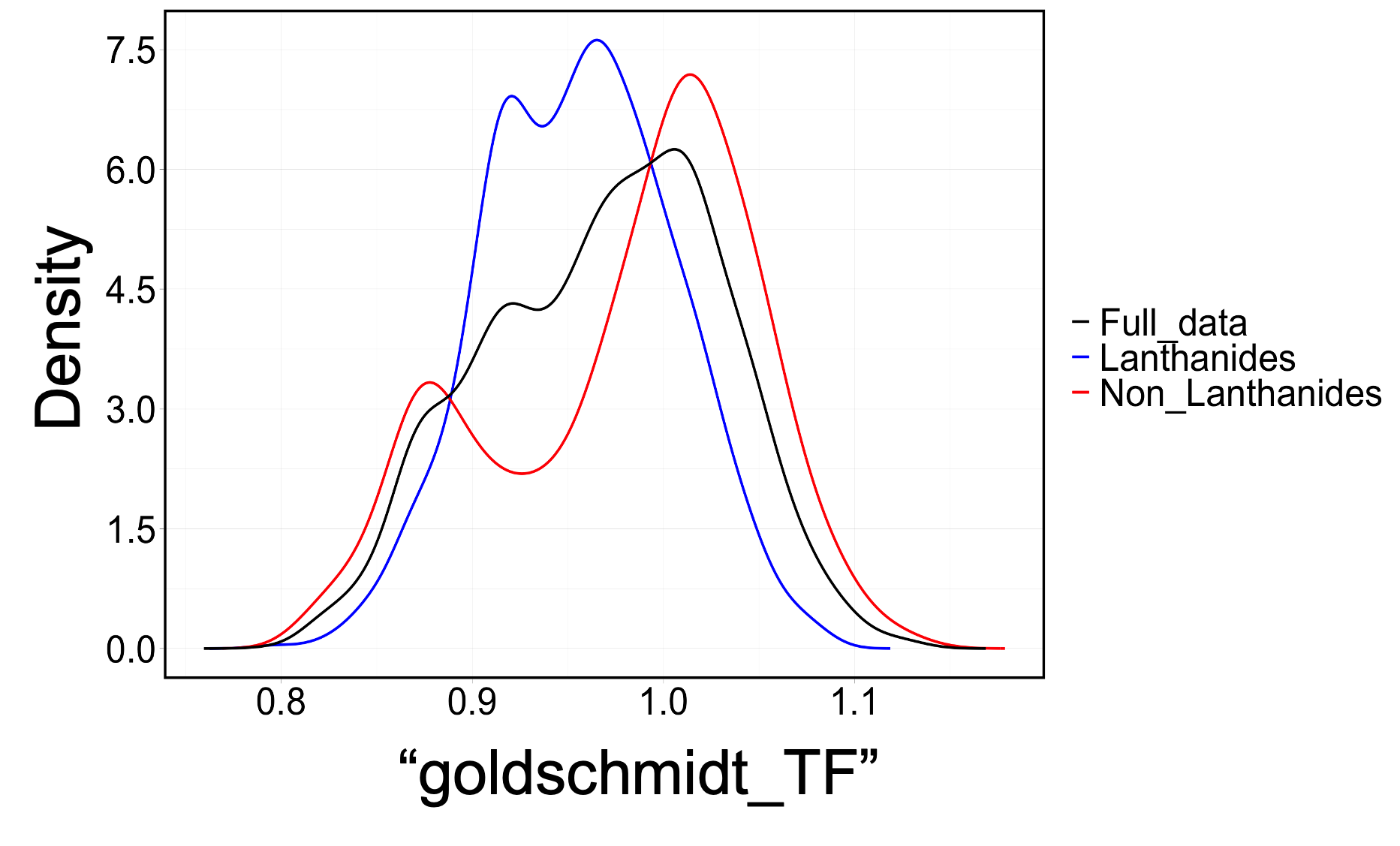}}
    \subfigure[p-value = $1.70\times10^{-9}$]{\label{Fig:goldion}\includegraphics[height=30 mm, width=45mm]{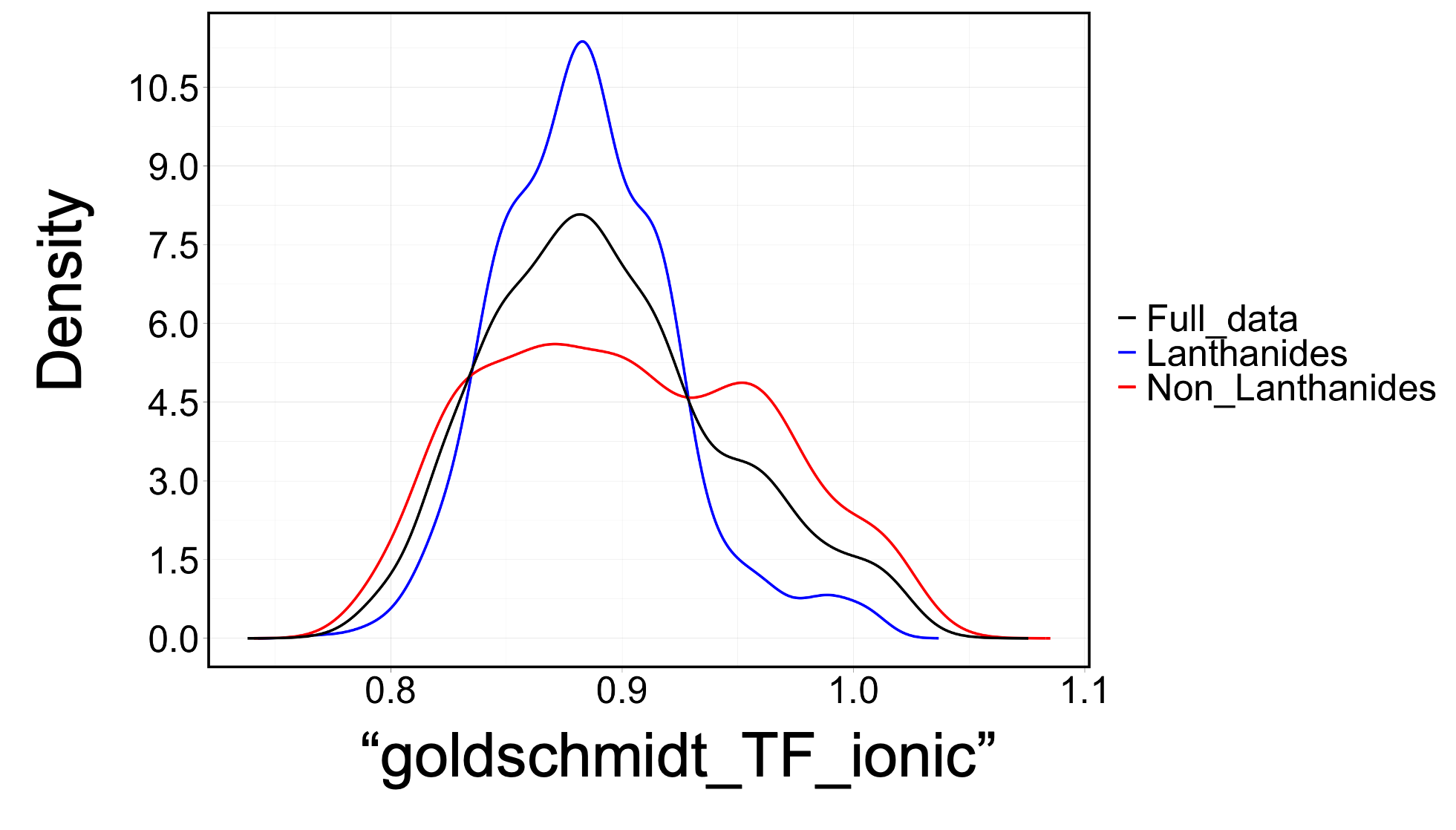}}
    \subfigure[p-value = $4.84\times10^{-1}$]{\label{Fig:oct}\includegraphics[height=30 mm, width=45mm]{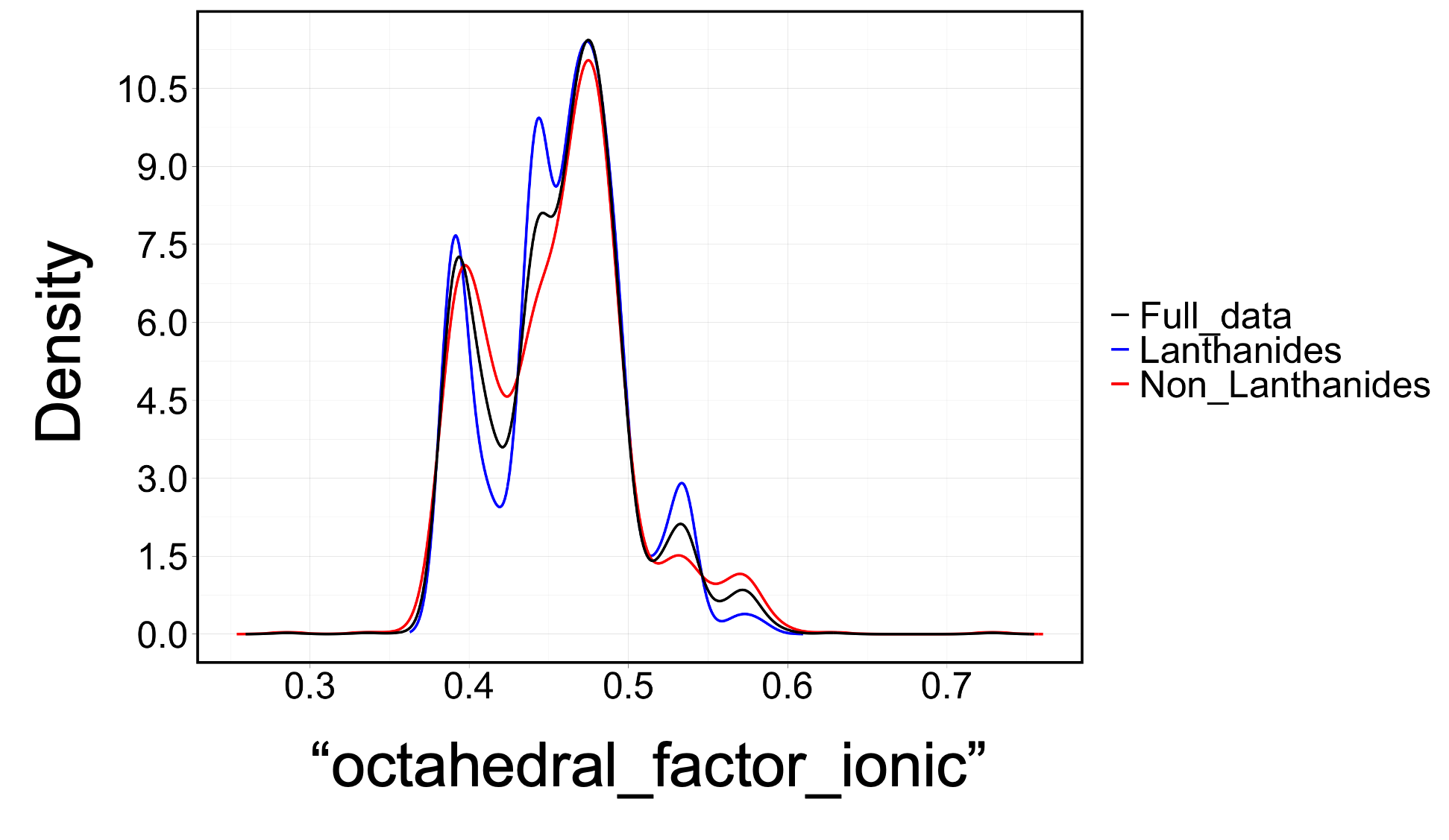}}
    \subfigure[p-value = $2.48\times10^{-1}$]
    {\label{Fig:oction}\includegraphics[height=30 mm, width=45mm]{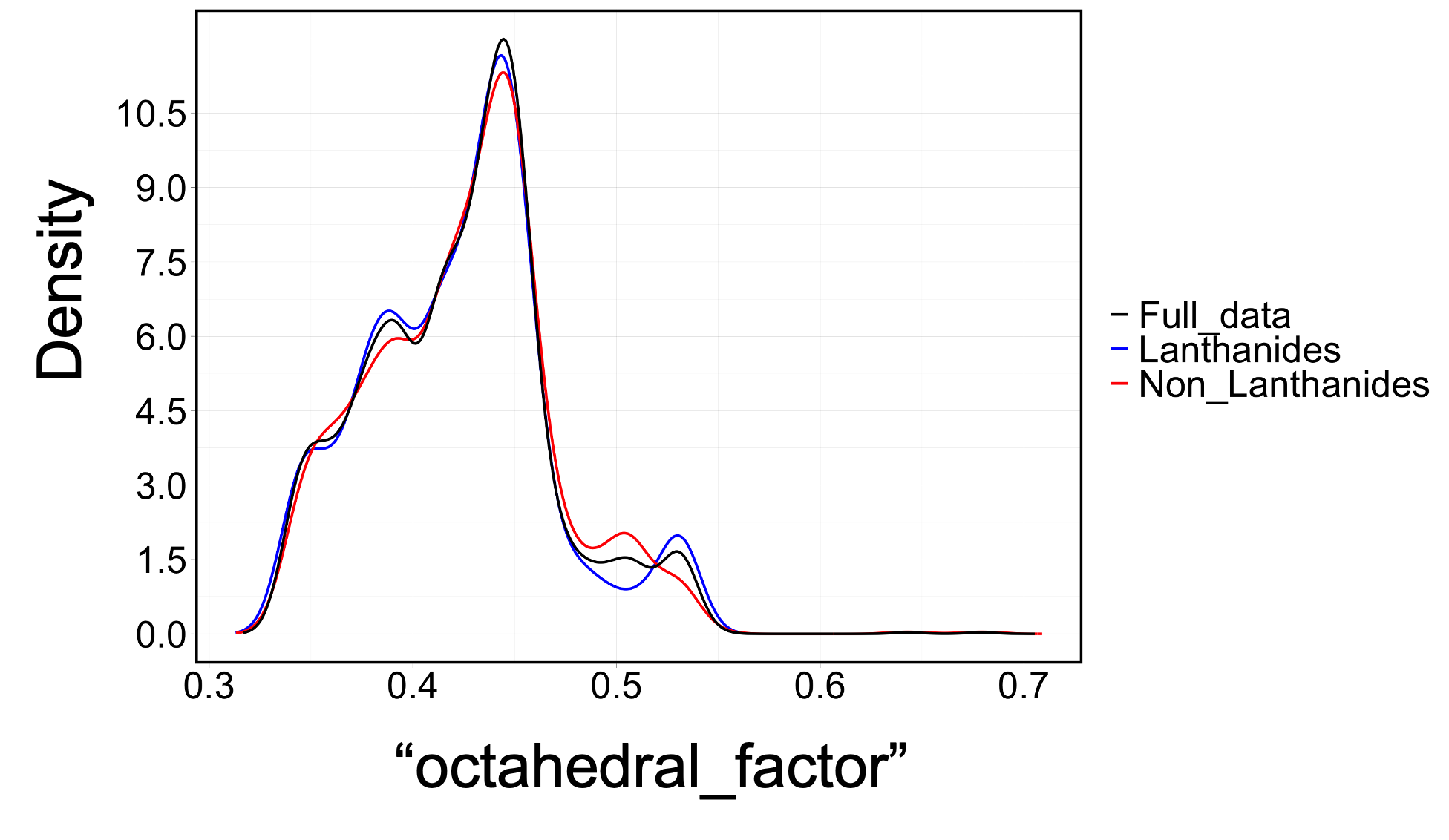}}
        \subfigure[p-value = $1.38\times10^{-25}$]{\label{Fig:AO}\includegraphics[height=30 mm, width=45mm]{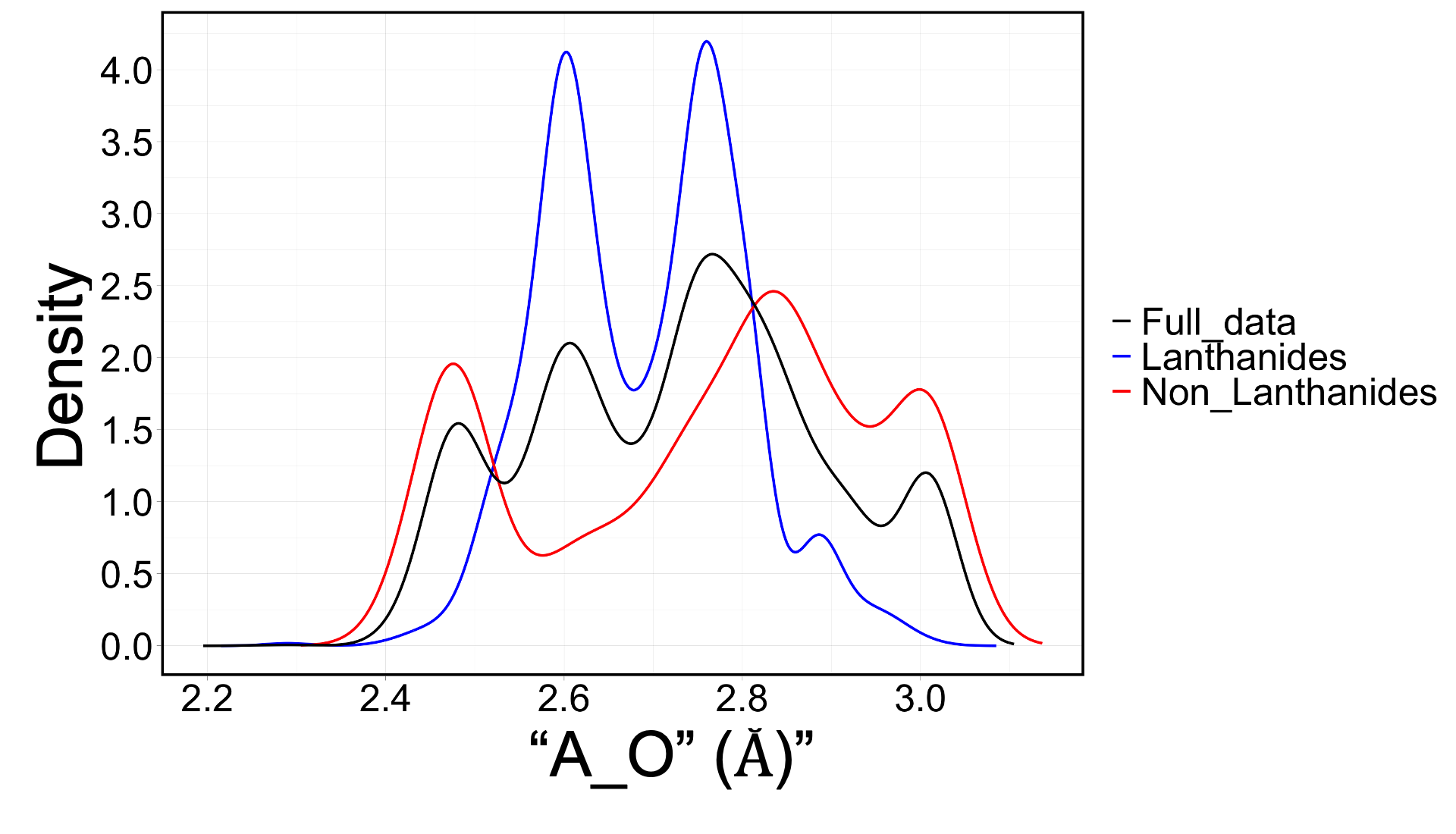}}
    \subfigure[p-value = $2.48\times10^{-1}$]{\label{Fig:BO}\includegraphics[height=30 mm, width=45mm]{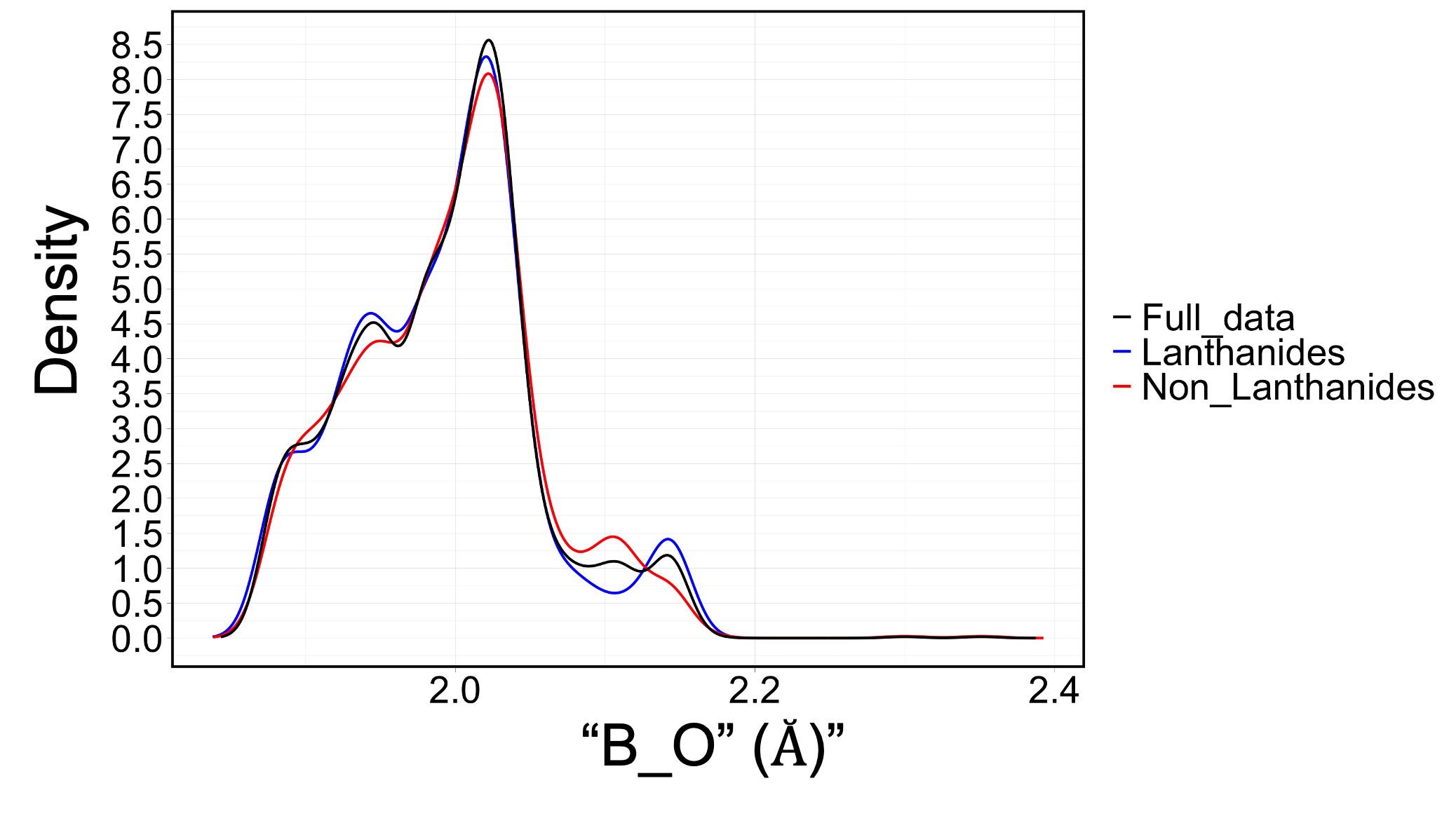}}
    \subfigure[p-value = $4.25\times10^{-19}$]{\label{Fig:AB}\includegraphics[height=30 mm, width=45mm]{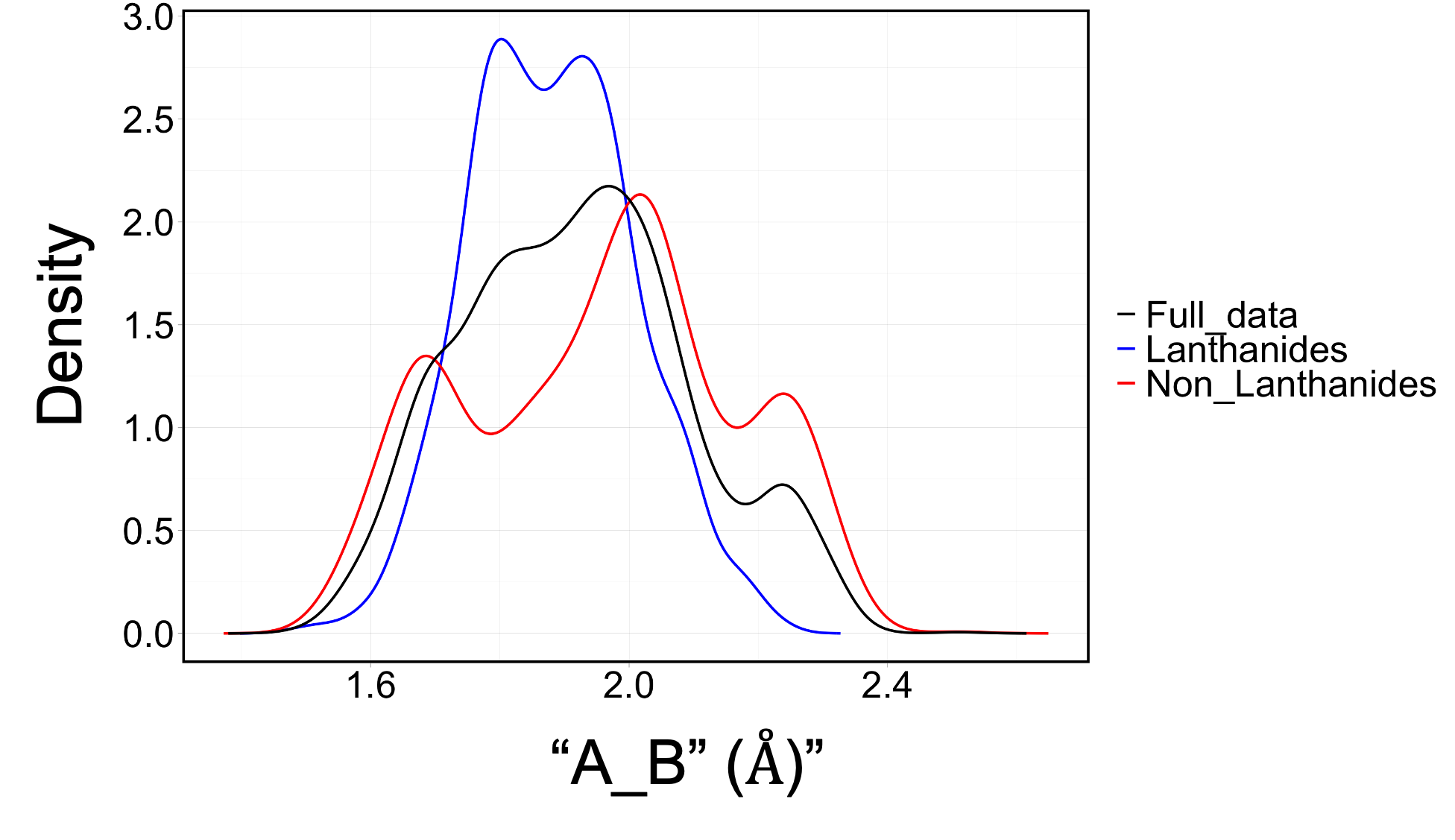}}
    \subfigure[p-value = $7.63\times10^{-33}$]{\label{Fig:hull}\includegraphics[height=30 mm, width=45mm]{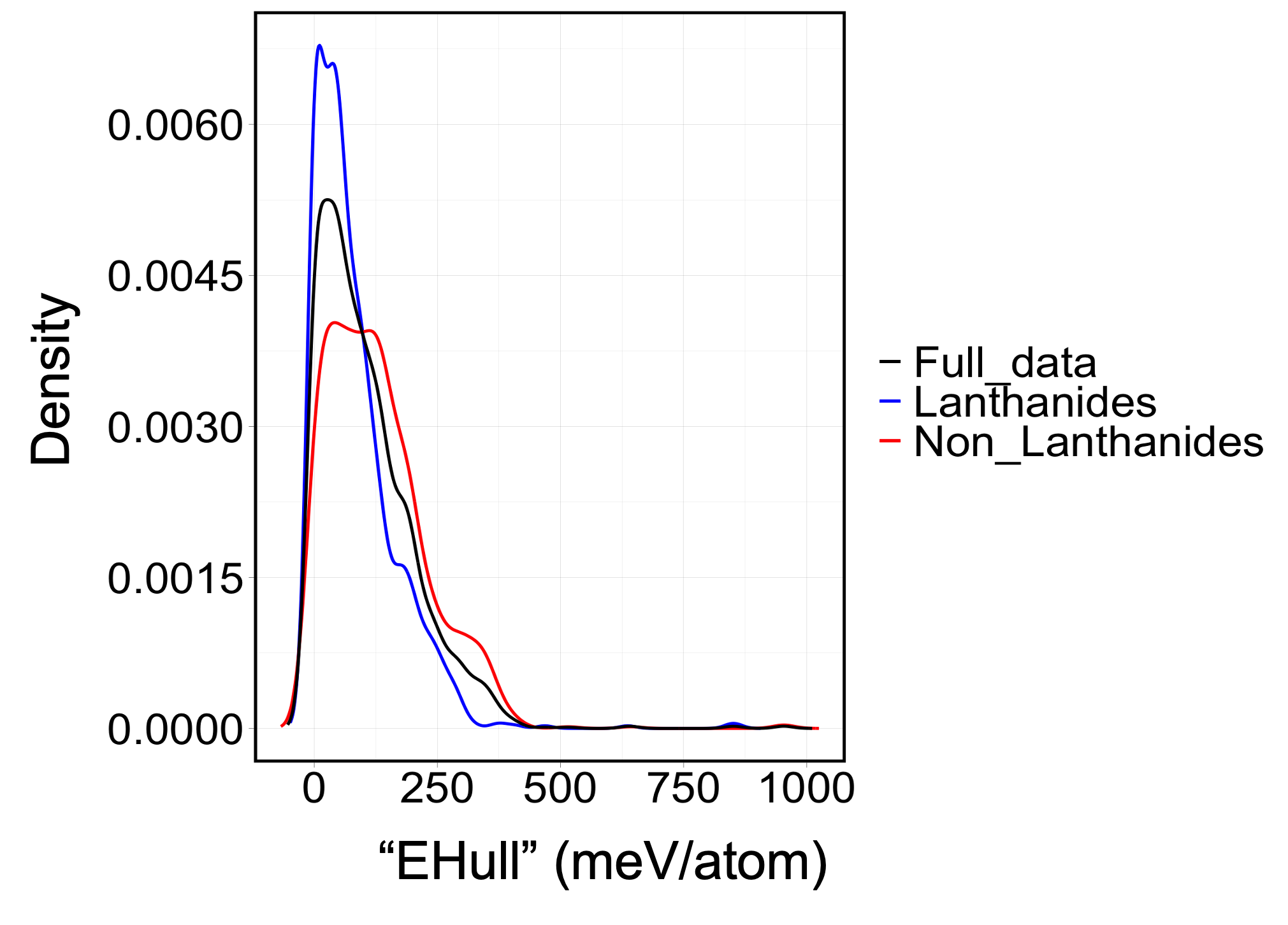}}
    \subfigure[p-value = $7.63\times10^{-33}$]{\label{Fig:form}\includegraphics[height=30 mm, width=45mm]{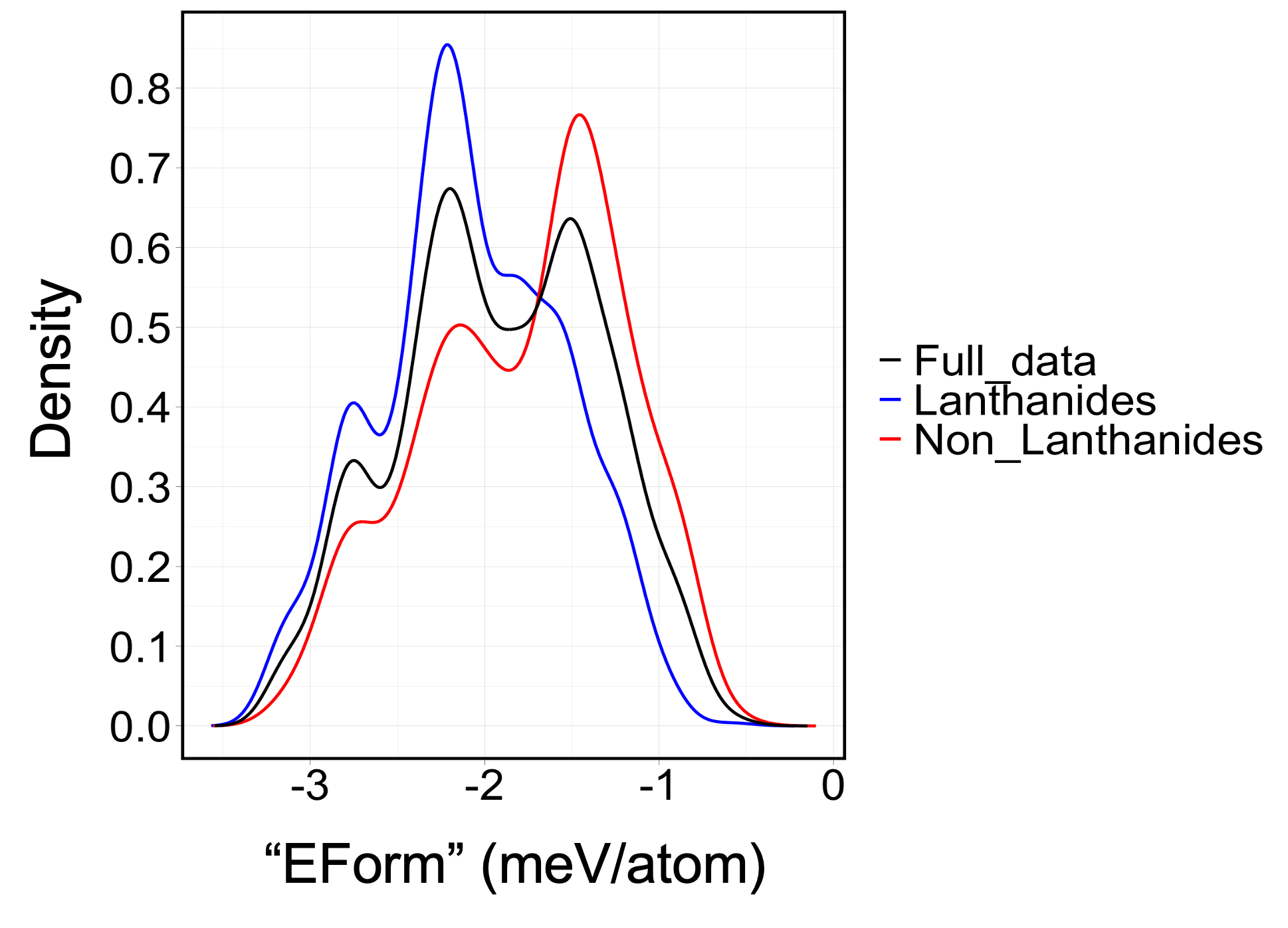}}

        \caption{Empirical probability density functions for lanthanum-containing substructures (shown in blue), non-lanthanum-containing structures (shown in red) and the entire data (shown in black) for (a)Goldschmidt tolerance factor, (b) Goldschmidt tolerance factor with ionic radii, 
        (c) octahedral factor with ionic radii, (d) octahedral factor,  (e) distance between A and oxide ions, (f) distance between B and oxide ions and (g) the average distance between A and B cations, respectively reported in {\AA} units. (h) EHull (i) EForm.
        For each case, we also compare lanthanum-containing and non-lanthanum-containing based data using nonparametric two-sample Wilcoxon tests and report the p-values (smaller p-value represents larger difference).}
            \label{fig:exploit_1}
    \end{figure}

 We observe differences in the estimated latent factors in our latent factor analysis for the top 100 screened features across the lanthanum-containing and non-lanthanum-containing substructures in Figures \ref{Fig:fa_l} and \ref{Fig:fa_nl}. Here we further show univariate empirical densities of the geometric blueprints for the above-mentioned substructures in Figure \ref{fig:exploit_1} and find that there are notable differences in the empirical densities of  Goldschmidt tolerance factor and its ionic counterpart. 
 Likewise, the densities given by the average distance between A-site and oxygen anion  (`A\_O') and A-site and B-site cation (`A\_B') are significantly different, while densities associated with octahedral factors and the distance between the B-site and the oxygen anion (`B\_O') remain constant across lanthanide-rich and non-lanthanide-rich data-fragments.
 In these figures, we also report the p-values associated with nonparametric two-sample Wilcoxon tests \citep{conover1999practical} to compare the respective lanthanum-containing and non-lanthanum-containing based data. A p-value smaller than 0.05 is typically considered, inferring the two groups to be statistically significantly different.
 Following this convention, we find that except for `octahedral\_factor', `octahedral\_factor\_ionic', and `B\_O', the rest have significant p-values, again highlighting the above-mentioned differences.

  In Table \ref{Table:3} we report the pooled mean of the out-of-sample MSEs, obtained by regressing geometric descriptors on EHull and EForm respectively. The pooled Mean is calculated following the working protocol mentioned in Figure \ref{Fig:flowchart}. In this analysis, we apply SVM, kernel ridge, Neural Network, and random forest. The hyperparameter optimization is done using the same protocol mentioned in Section \ref{sec: Analysis plan}. Out of the four reported errors, random forest registers the best fit for both of the two cases, with EHull and EForm as responses.   

\begin{figure}[htbp]
\label{Fig:lanthanide_regression}
    \centering
    \captionsetup{justification= centerlast}
   \subfigure[EHull $\sim$ Geometry]{\label{Fig:lanth0}\includegraphics[height=50 mm, width=75mm]{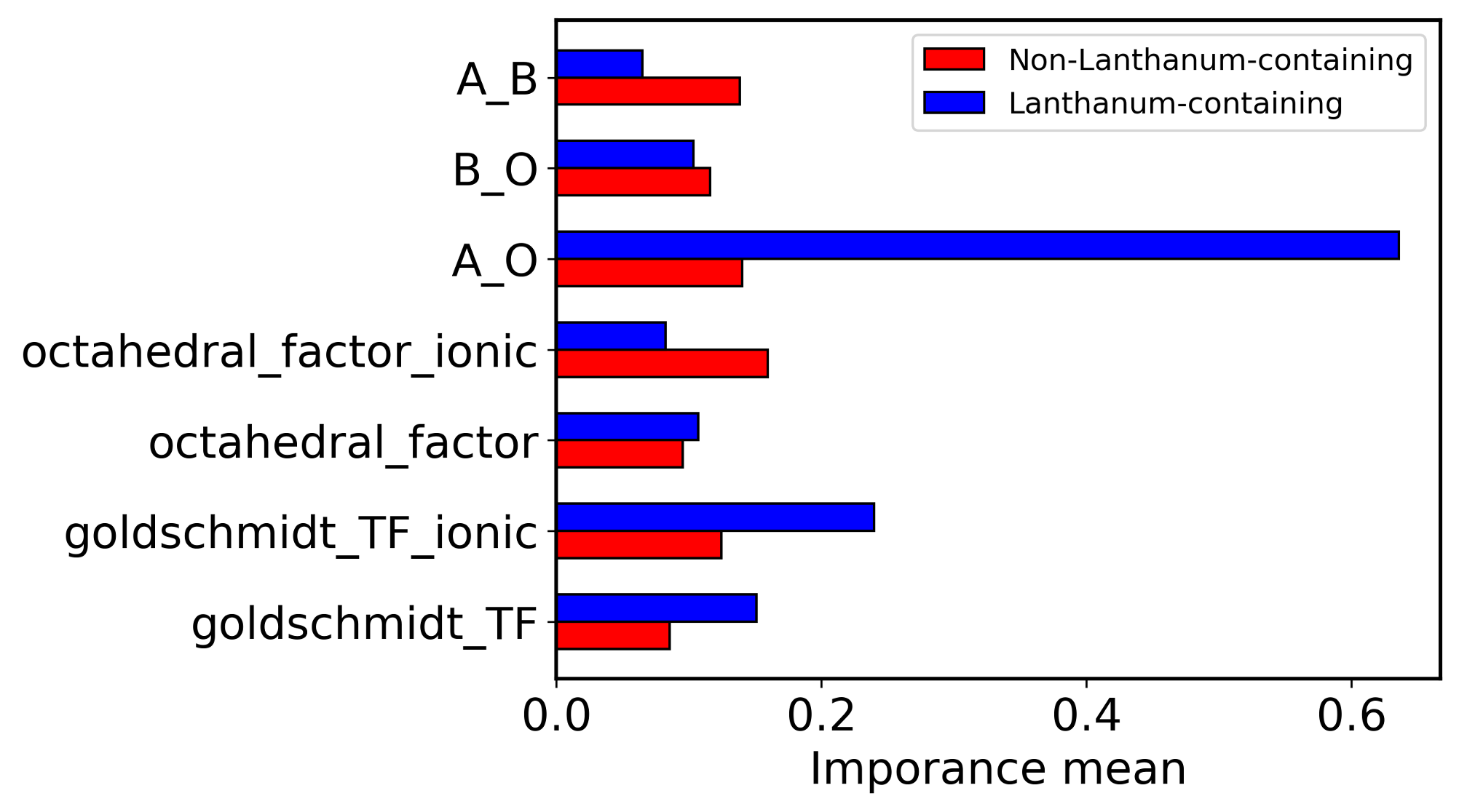}}
    \subfigure[EForm $\sim$ Geometry]{\label{Fig:lanth1}\includegraphics[height=50 mm, width=75mm]{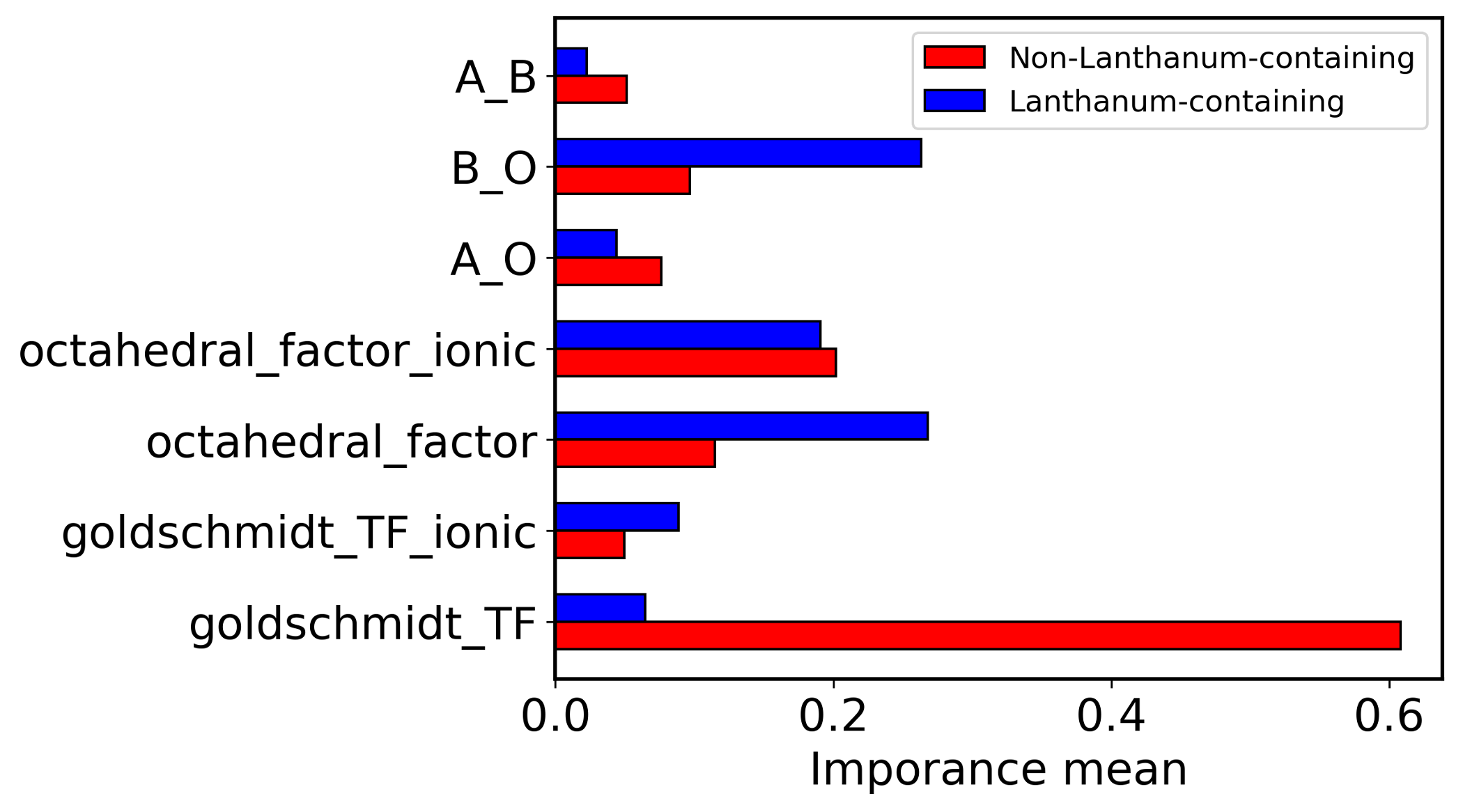}}
     \subfigure[EHull $\sim$ Geometry+Group indicator]{\label{Fig:lanth01}\includegraphics[height=50 mm, width=75mm]{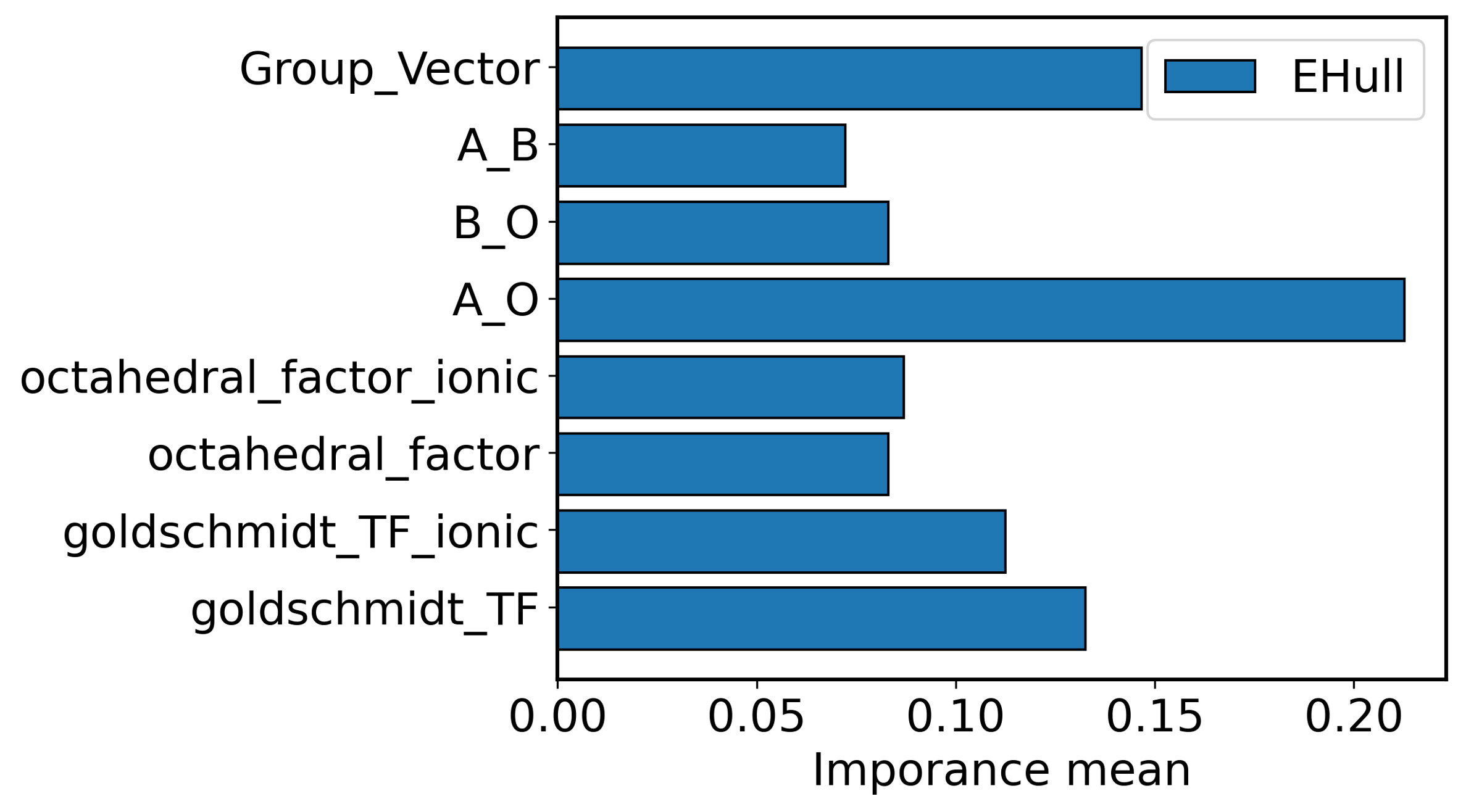}}
     \subfigure[EForm $\sim$ Geometry+Group indicator]{\label{Fig:lanth02}\includegraphics[height=50 mm, width=75mm]{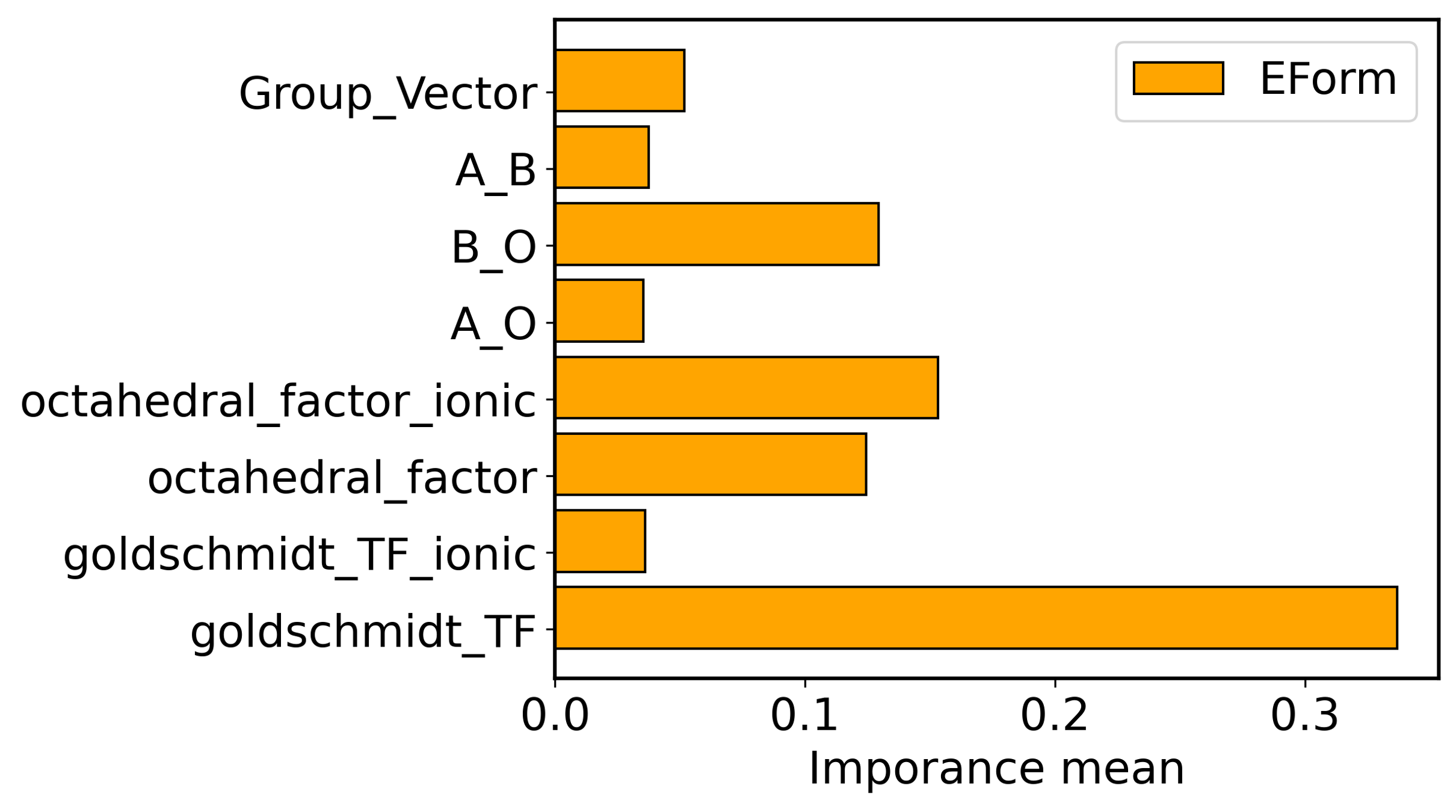}}
    \caption{Bar plots showing the important features for lanthanide  and non lanthanide compounds, regressing with (a) EHull as the response vector. (b) EForm as the response vector. X-axis show the selected features based on their importance mean on the  Y-axis. (c)-(d) Variable importance bar plot including the binary indicator vector for EHull and EForm respectively. The abbreviation description as, TF: Tolerance Factor, A\_O and B\_O: average distance between A/B cation and oxygen. A\_B: Average distance between A and B cations.}
 \end{figure}

Following the protocol in Figure~\ref{Fig:flowchart}, we identify the relevant features by applying permutation importance with the random forest as the regression model. Importance scores may not be directly comparable on an absolute scale. 
We thus analyze the relative ordering of the predictors within a given analysis based on their importance scores. As shown in Figures \ref{Fig:lanth0} and \ref{Fig:lanth1}, the relative orderings of the importance values vary across the lanthanum-containing and non-lanthanum-containing perovskites, as well as the choice of responses EHull and EForm. In Figure \ref{Fig:lanth0}, all the predictors seem to have relatively equal importance in explaining EHull for non-lanthanum-containing, but for the lanthanum-containing substructure, `A\_O' shows a much larger effect than all the other. However, in Figure \ref{Fig:lanth1} Goldshmidt tolerance factor is the most important for the non-lanthanum derived perovskites, while `B\_O' shows the largest relative importance for lanthanum-rich perovskites. This is an interesting finding. However, further research is needed, and possibly other datasets for a confirmatory understanding.
Nonetheless, `A\_O' is found to be the most important feature for all the cases, except for lanthanides with EForm as the response.

For a comprehensive study, we subsequently repeat our above analysis adding the lanthanide indicator vector as another predictor, and show the results in Figures \ref{Fig:lanth01} and \ref{Fig:lanth02}.
Although the indicator vector does not appear to be the most important, it still turns out to be the second most important for EHull (see Figure \ref{Fig:lanth01}) and also leads to a somewhat different order of importance of other predictors than the Figures \ref{Fig:lanth0} and \ref{Fig:lanth1}.
The distance between the A site and oxygen (`A\_O') is found to be the most significant factor impacting  EHull shown in Figure \ref{Fig:lanth0} . 
However, the importance of Goldschmidt tolerance factor  (`goldschmidt\_TF')  is significantly important, shown in Figure \ref{Fig:lanth1} in the presence of group vector while regressing with EForm. 


\begin{table}[htbp] 

\caption{Average out-of-sample MSEs over 20 randomiz  test-train splits of the data for the candidate regression models associated after regressing EHull and EForm energy values (meV/atom) on the seven geometric blueprints (the best performers are in bold).}
\label{Table:3}
\resizebox{0.55\textwidth}{!}{
\begin{tabular}{|c |c| c |}
\hline
Methods        & \begin{tabular}[c]{@{}c@{}}Pooled \\ Mean \\ MSE (EHull)\end{tabular} & \begin{tabular}[c]{@{}c@{}}Pooled \\ Mean \\ MSE (EForm)\end{tabular}  \\ \hline \hline
SVM            & 0.8782                                                               & 0.4620                                                                                                                                                                                              \\ \hline
Kernel Ridge   & 0.6387                                                              & 0.3935                                                                                                                                                                                             \\ \hline
Neural Network & 0.7656                                                                & 0.4432                                                                                                                                                                                                  \\ \hline
Random Forest  & {\bf 0.2106}                                                                & {\bf 0.4995}                                                                                                                                \\ \hline
\end{tabular}}
\end{table}

\vspace{-2mm}

\section{Conclusion}
\label{sec:conclusion}
Our machine learning analysis sheds light on the key structural features that govern \ce{ABO3} type oxide perovskites, as well as the extent of correlation with the stability and formability of these compounds. Since the regression is done using geometric blueprints as the response, it is reasonable to identify the descriptors explaining the geometric environment as the prominent features. For the latter part, the segregation of the perovskites on the basis of lanthanum-containing and non-lanthanum-containing changes the trend of the geometric feature vector, thereby providing a qualitative understanding of local structural effects such as bond elongation and octahedral rotation arising due to mixed lanthanide atoms in the structure. 
For our analyses, we applied different packages from {\tt scikit-learn} and R statistical software as outlined in Section~\ref{sec:computation}.
As in any high-dimensional data-driven analysis, we too apply a screening algorithm to first select a subset of the most relevant predictors.
Since the candidate regression models are mostly non-parametric, we consider a newly developed non-parametric screening method that also enjoys strong theoretical sure-screening properties \citep{xue2017robust}. In this work, the screened features are also shown to differ in their latent dependency patterns.
In our third analysis, we also identify several differential impacts of lanthanide-richness on the EHull and EForm energy profiles. 

It is true that geometric blueprints ($t$,$\mu$) often fail to capture the  static and dynamic rotation of octahedra associated with perovskite crystal structure \citep{jia2022dynamic}.  
As a result, using 1D geometric metric like $\tau$ \citep{bartel2019new}, using bond lengths and capturing any defects at A-site and B-site atoms from Scanning Electron Microscope (SEM) images \citep{choudhary2022recent} to depict octahedral environment with related oxygen vacancies are more quantitative approaches. SEM or experimental X-Ray Diffraction data (XRD) data would certainly enrich the dataset and aid in quantifying electrostatic effects \citep{hong2021local}  through a data driven approach, which we aim to address in our future work. 
Also, the utility of any data-driven method relies on the quality of the data. Our current work could further be improved by
utilizing electronic properties calculated with post-Hartree Fock methods \citep{suslick2001encyclopedia},
which theoretically 
yield results commensurate with experimental values.

Finally, our analysis in this paper leads to several interesting findings, motivating us to run some more follow-up analyses, which will be addressed in our future works. Firstly, we can run a mediation analysis \citep{alwin1975decomposition,mackinnon2012introduction} where the geometric features will be considered as potential mediators influencing the energy profiles after adjusting for the high dimensional structural descriptors.
Secondly, it will also be interesting to study differential characteristics in the high dimensional structural blueprints using high dimensional clustering methods across the lanthanum-containing and non-lanthanum-containing substructures.
To address these research directions, other datasets from \citep{jain2013commentary,curtarolo2012aflow, saal2013materials,ma2021discovery,emery2017high} could be beneficial.
The scope of this current work only focuses on the inference analysis with respect to the fundamental geometric metrics $t$ and $\mu$ and wishes to extend the analysis with data-driven descriptor $\tau$ \cite{bartel2019new} in our future works. 
\section*{Acknowledgement}
The first author would like to thank Prof. Yosuke Kanai for his encouragement and valuable comments on this work.
We would like to thank the editor, the associate editor, and two anonymous referees for their constructive suggestions that improved the quality of the manuscript.
\section*{Declaration of Competing Interest}
The authors declare that they have no known competing financial interests or personal relationships that could have appeared to
influence the work reported in this paper
\section*{Data Availability}
The data is publicly available in \cite{li2018data}. The associated codes and some additionally processed data are uploaded in \url{https://github.com/spriti523/Lanthanide-richness-vs-stability}. A Readme.docx file is also supplied with detailed implementation instructions and descriptions of the attached files to reproduce the results presented in this paper.
\bibliographystyle{plainnat}
\bibliography{bib}

\begin{thebibliography}{66}
\providecommand{\natexlab}[1]{#1}
\providecommand{\url}[1]{\texttt{#1}}
\expandafter\ifx\csname urlstyle\endcsname\relax
  \providecommand{\doi}[1]{doi: #1}\else
  \providecommand{\doi}{doi: \begingroup \urlstyle{rm}\Url}\fi

\bibitem[Aleksandrov and Bartolom{\'e}(2001)]{aleksandrov2001structural}
KS~Aleksandrov and J~Bartolom{\'e}.
\newblock Structural distortions in families of perovskite-like crystals.
\newblock \emph{Phase Transitions: A Multinational Journal}, 74\penalty0
  (3):\penalty0 255--335, 2001.

\bibitem[Alwin and Hauser(1975)]{alwin1975decomposition}
Duane~F Alwin and Robert~M Hauser.
\newblock The decomposition of effects in path analysis.
\newblock \emph{American sociological review}, pages 37--47, 1975.

\bibitem[Armiento et~al.(2014)Armiento, Kozinsky, Hautier, Fornari, and
  Ceder]{armiento2014high}
Rickard Armiento, Boris Kozinsky, Geoffroy Hautier, Marco Fornari, and Gerbrand
  Ceder.
\newblock High-throughput screening of perovskite alloys for piezoelectric
  performance and thermodynamic stability.
\newblock \emph{Physical Review B}, 89\penalty0 (13):\penalty0 134103, 2014.

\bibitem[Attfield(2002)]{attfield2002cation}
JP~Attfield.
\newblock ‘a’cation control of perovskite properties.
\newblock \emph{Crystal engineering}, 5\penalty0 (3-4):\penalty0 427--438,
  2002.

\bibitem[Bartel et~al.(2019)Bartel, Sutton, Goldsmith, Ouyang, Musgrave,
  Ghiringhelli, and Scheffler]{bartel2019new}
Christopher~J Bartel, Christopher Sutton, Bryan~R Goldsmith, Runhai Ouyang,
  Charles~B Musgrave, Luca~M Ghiringhelli, and Matthias Scheffler.
\newblock New tolerance factor to predict the stability of perovskite oxides
  and halides.
\newblock \emph{Science advances}, 5\penalty0 (2):\penalty0 eaav0693, 2019.

\bibitem[Belik et~al.(2016)Belik, Glazkova, Katsuya, Tanaka, Sobolev, and
  Presniakov]{belik2016low}
Alexei~A Belik, Yana~S Glazkova, Yoshio Katsuya, Masahiko Tanaka, Alexey~V
  Sobolev, and Igor~A Presniakov.
\newblock Low-temperature structural modulations in {CdMn7O12}, {CaMn7O12},
  {SrMn7O12}, and {PbMn7O12} perovskites studied by synchrotron x-ray powder
  diffraction and mossbauer spectroscopy.
\newblock \emph{The Journal of Physical Chemistry C}, 120\penalty0
  (15):\penalty0 8278--8288, 2016.

\bibitem[Bendersky et~al.(2003)Bendersky, Greenblatt, and
  Chen]{bendersky2003transmission}
Leonid~A Bendersky, Martha Greenblatt, and Rongji Chen.
\newblock Transmission electron microscopy study of ruddlesden--popper {Can+
  1MnnO3n+} 1 n= 2 and 3 compounds.
\newblock \emph{Journal of Solid State Chemistry}, 174\penalty0 (2):\penalty0
  418--423, 2003.

\bibitem[Breiman(2001)]{breiman2001random}
Leo Breiman.
\newblock Random forests.
\newblock \emph{Machine learning}, 45:\penalty0 5--32, 2001.

\bibitem[Bunte et~al.(2016)Bunte, Lepp{\"a}aho, Saarinen, and
  Kaski]{bunte2016sparse}
Kerstin Bunte, Eemeli Lepp{\"a}aho, Inka Saarinen, and Samuel Kaski.
\newblock Sparse group factor analysis for biclustering of multiple data
  sources.
\newblock \emph{Bioinformatics}, 32\penalty0 (16):\penalty0 2457--2463, 2016.

\bibitem[Cherif et~al.(2002)Cherif, Dhahri, Dhahri, Oumezzine, and
  Vincent]{cherif2002effect}
K~Cherif, J~Dhahri, E~Dhahri, M~Oumezzine, and H~Vincent.
\newblock Effect of the a cation size on the structural, magnetic, and
  electrical properties of perovskites {(La1- xNdx) 0.7 Sr0. 3r003nMnO3}.
\newblock \emph{Journal of Solid State Chemistry}, 163\penalty0 (2):\penalty0
  466--471, 2002.

\bibitem[Choudhary et~al.(2022)Choudhary, DeCost, Chen, Jain, Tavazza, Cohn,
  Park, Choudhary, Agrawal, Billinge, et~al.]{choudhary2022recent}
Kamal Choudhary, Brian DeCost, Chi Chen, Anubhav Jain, Francesca Tavazza, Ryan
  Cohn, Cheol~Woo Park, Alok Choudhary, Ankit Agrawal, Simon~JL Billinge,
  et~al.
\newblock Recent advances and applications of deep learning methods in
  materials science.
\newblock \emph{npj Computational Materials}, 8\penalty0 (1):\penalty0 59,
  2022.

\bibitem[Conover(1999)]{conover1999practical}
William~Jay Conover.
\newblock \emph{Practical nonparametric statistics}, volume 350.
\newblock john wiley \& sons, 1999.

\bibitem[Curtarolo et~al.(2012)Curtarolo, Setyawan, Hart, Jahnatek, Chepulskii,
  Taylor, Wang, Xue, Yang, Levy, et~al.]{curtarolo2012aflow}
Stefano Curtarolo, Wahyu Setyawan, Gus~LW Hart, Michal Jahnatek, Roman~V
  Chepulskii, Richard~H Taylor, Shidong Wang, Junkai Xue, Kesong Yang, Ohad
  Levy, et~al.
\newblock Aflow: An automatic framework for high-throughput materials
  discovery.
\newblock \emph{Computational Materials Science}, 58:\penalty0 218--226, 2012.

\bibitem[Emery and Wolverton(2017)]{emery2017high}
Antoine~A Emery and Chris Wolverton.
\newblock High-throughput dft calculations of formation energy, stability and
  oxygen vacancy formation energy of abo3 perovskites.
\newblock \emph{Scientific data}, 4\penalty0 (1):\penalty0 1--10, 2017.

\bibitem[Fan and Lv(2008)]{fan2008sure}
Jianqing Fan and Jinchi Lv.
\newblock Sure independence screening for ultrahigh dimensional feature space.
\newblock \emph{Journal of the Royal Statistical Society Series B: Statistical
  Methodology}, 70\penalty0 (5):\penalty0 849--911, 2008.

\bibitem[Ferbinteanu et~al.(2017)Ferbinteanu, Stroppa, Scarrozza, Humelnicu,
  Maftei, Frecus, and Cimpoesu]{ferbinteanu2017density}
Marilena Ferbinteanu, Alessandro Stroppa, Marco Scarrozza, Ionel Humelnicu, Dan
  Maftei, Bogdan Frecus, and Fanica Cimpoesu.
\newblock On the density functional theory treatment of lanthanide coordination
  compounds: a comparative study in a series of cu--ln (ln= gd, tb, lu)
  binuclear complexes.
\newblock \emph{Inorganic Chemistry}, 56\penalty0 (16):\penalty0 9474--9485,
  2017.

\bibitem[Filip and Giustino(2018)]{filip2018geometric}
Marina~R Filip and Feliciano Giustino.
\newblock The geometric blueprint of perovskites.
\newblock \emph{Proceedings of the National Academy of Sciences}, 115\penalty0
  (21):\penalty0 5397--5402, 2018.

\bibitem[G{\'e}ron(2017)]{geron2017hands}
Aur{\'e}lien G{\'e}ron.
\newblock Hands-on machine learning with scikit-learn and tensorflow: Concepts.
\newblock \emph{Tools, and Techniques to build intelligent systems}, 2017.

\bibitem[Giaquinta and Zur~Loye(1994)]{giaquinta1994structural}
Daniel~M Giaquinta and Hans-Conrad Zur~Loye.
\newblock Structural predictions in the abo3 phase diagram.
\newblock \emph{Chemistry of materials}, 6\penalty0 (4):\penalty0 365--372,
  1994.

\bibitem[Glazer and Megaw(1972)]{glazer1972structure}
AM~Glazer and Helen~D Megaw.
\newblock The structure of sodium niobate (t2) at 600° c, and the
  cubic-tetragonal transition in relation to soft-phonon modes.
\newblock \emph{Philosophical Magazine}, 25\penalty0 (5):\penalty0 1119--1135,
  1972.

\bibitem[Goldschmidt(1926)]{goldschmidt1926gesetze}
Victor~Moritz Goldschmidt.
\newblock Die gesetze der krystallochemie.
\newblock \emph{Naturwissenschaften}, 14\penalty0 (21):\penalty0 477--485,
  1926.

\bibitem[Gopalakrishnan et~al.(2020)Gopalakrishnan, Sebastian, and
  Ahn]{gopalakrishnan2020perovskite}
Pratheek Gopalakrishnan, Ann~Rose Sebastian, and Ethan~C Ahn.
\newblock Perovskite oxides tunable by electromechanical and electrothermal
  couplings.
\newblock \emph{ECS Transactions}, 98\penalty0 (3):\penalty0 87, 2020.

\bibitem[Goudochnikov and Bell(2007)]{goudochnikov2007correlations}
Pavel Goudochnikov and Andrew~J Bell.
\newblock Correlations between transition temperature, tolerance factor and
  cohesive energy in 2+: 4+ perovskites.
\newblock \emph{Journal of Physics: Condensed Matter}, 19\penalty0
  (17):\penalty0 176201, 2007.

\bibitem[Hayward et~al.(2002)Hayward, Cussen, Claridge, Bieringer, Rosseinsky,
  Kiely, Blundell, Marshall, and Pratt]{hayward2002hydride}
MA~Hayward, EJ~Cussen, JB~Claridge, M~Bieringer, MJ~Rosseinsky, CJ~Kiely,
  SJ~Blundell, IM~Marshall, and FL~Pratt.
\newblock The hydride anion in an extended transition metal oxide array:
  {LaSrCoO3H0}. 7.
\newblock \emph{Science}, 295\penalty0 (5561):\penalty0 1882--1884, 2002.

\bibitem[Henze and Zirkler(1990)]{henze1990class}
Norbert Henze and B~Zirkler.
\newblock A class of invariant consistent tests for multivariate normality.
\newblock \emph{Communications in statistics-Theory and Methods}, 19\penalty0
  (10):\penalty0 3595--3617, 1990.

\bibitem[Hilpert et~al.(2003)Hilpert, Steinbrech, Boroomand, Wessel, Meschke,
  Zuev, Teller, Nickel, and Singheiser]{hilpert2003defect}
K~Hilpert, RW~Steinbrech, F~Boroomand, E~Wessel, F~Meschke, A~Zuev, O~Teller,
  H~Nickel, and L~Singheiser.
\newblock Defect formation and mechanical stability of perovskites based on
  {LaCrO3} for solid oxide fuel cells {(SOFC)}.
\newblock \emph{Journal of the European Ceramic Society}, 23\penalty0
  (16):\penalty0 3009--3020, 2003.

\bibitem[Hong et~al.(2021)Hong, Byeon, Bak, Heo, Kim, Bae, and
  Chung]{hong2021local}
Youngjae Hong, Pilgyu Byeon, Jumi Bak, Yoon Heo, Hye-Sung Kim, Hyung~Bin Bae,
  and Sung-Yoon Chung.
\newblock Local-electrostatics-induced oxygen octahedral distortion in
  perovskite oxides and insight into the structure of ruddlesden--popper
  phases.
\newblock \emph{Nature communications}, 12\penalty0 (1):\penalty0 1--10, 2021.

\bibitem[Jacobs et~al.(2018)Jacobs, Mayeshiba, Booske, and
  Morgan]{jacobs2018material}
Ryan Jacobs, Tam Mayeshiba, John Booske, and Dane Morgan.
\newblock Material discovery and design principles for stable, high activity
  perovskite cathodes for solid oxide fuel cells.
\newblock \emph{Advanced Energy Materials}, 8\penalty0 (11):\penalty0 1702708,
  2018.

\bibitem[Jain et~al.(2013)Jain, Ong, Hautier, Chen, Richards, Dacek, Cholia,
  Gunter, Skinner, Ceder, et~al.]{jain2013commentary}
Anubhav Jain, Shyue~Ping Ong, Geoffroy Hautier, Wei Chen, William~Davidson
  Richards, Stephen Dacek, Shreyas Cholia, Dan Gunter, David Skinner, Gerbrand
  Ceder, et~al.
\newblock Commentary: The materials project: A materials genome approach to
  accelerating materials innovation.
\newblock \emph{APL materials}, 1\penalty0 (1):\penalty0 011002, 2013.

\bibitem[Jia et~al.(2020)Jia, Hu, Xu, Gao, Zhao, Barone, Stroppa, and
  Ren]{jia2020persistent}
Fanhao Jia, Shunbo Hu, Shaowen Xu, Heng Gao, Guodong Zhao, Paolo Barone,
  Alessandro Stroppa, and Wei Ren.
\newblock Persistent spin-texture and ferroelectric polarization in 2d hybrid
  perovskite benzylammonium lead-halide.
\newblock \emph{The journal of physical chemistry letters}, 11\penalty0
  (13):\penalty0 5177--5183, 2020.

\bibitem[Jia et~al.(2022)Jia, He, Akhtar, Herranz, and Pruneda]{jia2022dynamic}
Jiahui Jia, Xu~He, Arsalan Akhtar, Gervasi Herranz, and Miguel Pruneda.
\newblock Dynamic control of octahedral rotation in perovskites by defect
  engineering.
\newblock \emph{Physical Review B}, 105\penalty0 (22):\penalty0 224112, 2022.

\bibitem[Jin et~al.(2008)Jin, Zhou, Goodenough, Liu, Zhao, Yang, Yu, Yu,
  Katsura, Shatskiy, et~al.]{jin2008high}
C-Q Jin, J-S Zhou, JB~Goodenough, QQ~Liu, JG~Zhao, LX~Yang, Y~Yu, RC~Yu,
  T~Katsura, A~Shatskiy, et~al.
\newblock High-pressure synthesis of the cubic perovskite {BaRuO3} and
  evolution of ferromagnetism in {ARuO3 (A= Ca, Sr, Ba)} ruthenates.
\newblock \emph{Proceedings of the National Academy of Sciences}, 105\penalty0
  (20):\penalty0 7115--7119, 2008.

\bibitem[Johnsson and Lemmens(2007)]{johnsson2005crystallography}
Mats Johnsson and Peter Lemmens.
\newblock \emph{Crystallography and Chemistry of Perovskites}.
\newblock John Wiley \& Sons, Ltd, 2007.
\newblock ISBN 9780470022184.

\bibitem[Li et~al.(2022)Li, Lin, Liu, Hu, Cao, Chen, and Xing]{li2022chemical}
Qiang Li, Kun Lin, Zhanning Liu, Lei Hu, Yili Cao, Jun Chen, and Xianran Xing.
\newblock Chemical diversity for tailoring negative thermal expansion.
\newblock \emph{Chemical Reviews}, 122\penalty0 (9):\penalty0 8438--8486, 2022.

\bibitem[Li et~al.(2017)Li, Wang, Deschler, Gao, Friend, and
  Cheetham]{li2017chemically}
Wei Li, Zheming Wang, Felix Deschler, Song Gao, Richard~H Friend, and Anthony~K
  Cheetham.
\newblock Chemically diverse and multifunctional hybrid organic--inorganic
  perovskites.
\newblock \emph{Nature Reviews Materials}, 2\penalty0 (3):\penalty0 1--18,
  2017.

\bibitem[Li et~al.(2018{\natexlab{a}})Li, Jacobs, and Morgan]{li2018data}
Wei Li, Ryan Jacobs, and Dane Morgan.
\newblock Data and supplemental information for predicting the thermodynamic
  stability of perovskite oxides using machine learning models.
\newblock \emph{Data in brief}, 19:\penalty0 261--263, 2018{\natexlab{a}}.

\bibitem[Li et~al.(2018{\natexlab{b}})Li, Jacobs, and Morgan]{li2018predicting}
Wei Li, Ryan Jacobs, and Dane Morgan.
\newblock Predicting the thermodynamic stability of perovskite oxides using
  machine learning models.
\newblock \emph{Computational Materials Science}, 150:\penalty0 454--463,
  2018{\natexlab{b}}.

\bibitem[Li et~al.(2015)Li, Wang, He, Liu, Long, Han, and Pan]{li2015high}
Xiuzhi Li, Zujian Wang, Chao He, Ying Liu, Xifa Long, Shujuan Han, and Shilie
  Pan.
\newblock High piezoelectric response of a new ternary ferroelectric {Pb
  (Ho1/2Nb1/2) O3-Pb (Mg1/3Nb2/3) O3-PbTiO3 single crystal}.
\newblock \emph{Materials Letters}, 143:\penalty0 88--90, 2015.

\bibitem[Liang et~al.(2020)Liang, Lin, Lan, Meng, Zhao, Zou, Castelli,
  Pullerits, Canton, and Zheng]{liang2020electronic}
Mingli Liang, Weihua Lin, Zhenyun Lan, Jie Meng, Qian Zhao, Xianshao Zou,
  Ivano~E Castelli, Tonu Pullerits, Sophie~E Canton, and Kaibo Zheng.
\newblock Electronic structure and trap states of two-dimensional
  ruddlesden--popper perovskites with the relaxed goldschmidt tolerance factor.
\newblock \emph{ACS Applied Electronic Materials}, 2\penalty0 (5):\penalty0
  1402--1412, 2020.

\bibitem[Liang et~al.(2008)Liang, Tang, Shao, Li, Zeng, and
  Zheng]{liang2008synthesis}
Zhenhua Liang, Kaibin Tang, Qian Shao, Guocan Li, Suyuan Zeng, and Huagui
  Zheng.
\newblock Synthesis, crystal structure, and photocatalytic activity of a new
  two-layer ruddlesden--popper phase, {Li2CaTa2O7}.
\newblock \emph{Journal of Solid State Chemistry}, 181\penalty0 (4):\penalty0
  964--970, 2008.

\bibitem[Liu et~al.(2020)Liu, Cheng, Dong, Feng, Pang, Tian, Ma, Xia, Zhang,
  and Dong]{liu2020screening}
Haiying Liu, Jiucheng Cheng, Hongzhou Dong, Jianguang Feng, Beili Pang, Ziya
  Tian, Shuai Ma, Fengjin Xia, Chunkai Zhang, and Lifeng Dong.
\newblock Screening stable and metastable abo3 perovskites using machine
  learning and the materials project.
\newblock \emph{Computational Materials Science}, 177:\penalty0 109614, 2020.

\bibitem[Liu et~al.(2009)Liu, Lafferty, and Wasserman]{liu2009nonparanormal}
Han Liu, John Lafferty, and Larry Wasserman.
\newblock The nonparanormal: Semiparametric estimation of high dimensional
  undirected graphs.
\newblock \emph{Journal of Machine Learning Research}, 10\penalty0 (10), 2009.

\bibitem[Liu et~al.(2015)Liu, Rong, Malik, Canepa, Jain, Ceder, and
  Persson]{liu2015spinel}
Miao Liu, Ziqin Rong, Rahul Malik, Pieremanuele Canepa, Anubhav Jain, Gerbrand
  Ceder, and Kristin~A Persson.
\newblock Spinel compounds as multivalent battery cathodes: a systematic
  evaluation based on ab initio calculations.
\newblock \emph{Energy \& Environmental Science}, 8\penalty0 (3):\penalty0
  964--974, 2015.

\bibitem[Ma et~al.(2021)Ma, Jacobs, Booske, and Morgan]{ma2021discovery}
Tianyu Ma, Ryan Jacobs, John Booske, and Dane Morgan.
\newblock Discovery and engineering of low work function perovskite materials.
\newblock \emph{Journal of Materials Chemistry C}, 9\penalty0 (37):\penalty0
  12778--12790, 2021.

\bibitem[MacKinnon(2012)]{mackinnon2012introduction}
David~P MacKinnon.
\newblock \emph{Introduction to statistical mediation analysis}.
\newblock Routledge, 2012.

\bibitem[McCulloch and Pitts(1943)]{mcculloch1943logical}
Warren~S McCulloch and Walter Pitts.
\newblock A logical calculus of the ideas immanent in nervous activity.
\newblock \emph{The bulletin of mathematical biophysics}, 5:\penalty0 115--133,
  1943.

\bibitem[Nag and Shubha(2014)]{nag2014oxide}
Abanti Nag and V~Shubha.
\newblock Oxide thermoelectric materials: A structure--property relationship.
\newblock \emph{Journal of electronic materials}, 43:\penalty0 962--977, 2014.

\bibitem[Pedregosa et~al.(2011)Pedregosa, Varoquaux, Gramfort, Michel, Thirion,
  Grisel, Blondel, Prettenhofer, Weiss, Dubourg, Vanderplas, Passos,
  Cournapeau, Brucher, Perrot, and Duchesnay]{scikit-learn}
F.~Pedregosa, G.~Varoquaux, A.~Gramfort, V.~Michel, B.~Thirion, O.~Grisel,
  M.~Blondel, P.~Prettenhofer, R.~Weiss, V.~Dubourg, J.~Vanderplas, A.~Passos,
  D.~Cournapeau, M.~Brucher, M.~Perrot, and E.~Duchesnay.
\newblock Scikit-learn: Machine learning in {P}ython.
\newblock \emph{Journal of Machine Learning Research}, 12:\penalty0 2825--2830,
  2011.

\bibitem[Refaeilzadeh et~al.(2009)Refaeilzadeh, Tang, Liu,
  et~al.]{refaeilzadeh2009cross}
Payam Refaeilzadeh, Lei Tang, Huan Liu, et~al.
\newblock Cross-validation.
\newblock \emph{Encyclopedia of database systems}, 5:\penalty0 532--538, 2009.

\bibitem[Richter et~al.(2009)Richter, Holtappels, Graule, Nakamura, and
  Gauckler]{richter2009materials}
J{\"o}rg Richter, Peter Holtappels, Thomas Graule, Tetsuro Nakamura, and
  Ludwig~J Gauckler.
\newblock Materials design for perovskite {SOFC} cathodes.
\newblock \emph{Monatshefte f{\"u}r Chemie-Chemical Monthly}, 140\penalty0
  (9):\penalty0 985--999, 2009.

\bibitem[Saal et~al.(2013)Saal, Kirklin, Aykol, Meredig, and
  Wolverton]{saal2013materials}
James~E Saal, Scott Kirklin, Muratahan Aykol, Bryce Meredig, and Christopher
  Wolverton.
\newblock Materials design and discovery with high-throughput density
  functional theory: the open quantum materials database {(OQMD)}.
\newblock \emph{Jom}, 65:\penalty0 1501--1509, 2013.

\bibitem[Schader et~al.(2017)Schader, Rossetti~Jr, Luo, and
  Webber]{schader2017piezoelectric}
Florian~H Schader, George~A Rossetti~Jr, Jun Luo, and Kyle~G Webber.
\newblock Piezoelectric and ferroelectric properties of< 001> {C Pb
  (In1/2Nb1/2) O3-Pb (Mg1/3Nb2/3) O3-PbTiO3} single crystals under combined
  thermal and mechanical loading.
\newblock \emph{Acta Materialia}, 126:\penalty0 174--181, 2017.

\bibitem[Shetty et~al.(2022)Shetty, Shedthi, and
  Kumaraswamy]{shetty2022predicting}
Vidyasagar Shetty, Shabari Shedthi, and J~Kumaraswamy.
\newblock Predicting the thermodynamic stability of perovskite oxides using
  multiple machine learning techniques.
\newblock \emph{Materials Today: Proceedings}, 52:\penalty0 457--461, 2022.

\bibitem[Song et~al.(2018)Song, Li, Guo, Xu, and Fan]{song2018compositional}
Kexin Song, Zhenrong Li, Haisheng Guo, Zhuo Xu, and Shiji Fan.
\newblock Compositional segregation and electrical properties characterization
  of [001]-and [011]-oriented co-growth {Pb (In1/2Nb1/2) O3-Pb (Mg1/3Nb2/3)
  O3-PbTiO3} single crystal.
\newblock \emph{Journal of Applied Physics}, 123\penalty0 (15):\penalty0
  154107, 2018.

\bibitem[Sun et~al.(2020)Sun, Dougherty, Huang, Li, and
  Yan]{sun2020accelerating}
Mingzi Sun, Alan~William Dougherty, Bolong Huang, Yuliang Li, and Chun-Hua Yan.
\newblock Accelerating atomic catalyst discovery by theoretical
  calculations-machine learning strategy.
\newblock \emph{Advanced Energy Materials}, 10\penalty0 (12):\penalty0 1903949,
  2020.

\bibitem[Suslick(2001)]{suslick2001encyclopedia}
Kenneth~S Suslick.
\newblock Encyclopedia of physical science and technology.
\newblock \emph{Sonoluminescence and sonochemistry, 3rd edn. Elsevier Science
  Ltd, Massachusetts}, pages 1--20, 2001.

\bibitem[Vapnik(1999)]{vapnik1999nature}
Vladimir Vapnik.
\newblock \emph{The nature of statistical learning theory}.
\newblock Springer science \& business media, 1999.

\bibitem[Wang et~al.(2016)Wang, He, Li, Liu, Long, Han, and
  Pan]{wang2016scandium}
Zujian Wang, Chao He, Xiuzhi Li, Ying Liu, Xifa Long, Shujuan Han, and Shilie
  Pan.
\newblock Scandium modified lead magnesium niobate-lead titanate single
  crystals for high temperature and high power applications.
\newblock \emph{Materials Letters}, 184:\penalty0 162--165, 2016.

\bibitem[Ward et~al.(2016)Ward, Agrawal, Choudhary, and
  Wolverton]{ward2016general}
Logan Ward, Ankit Agrawal, Alok Choudhary, and Christopher Wolverton.
\newblock A general-purpose machine learning framework for predicting
  properties of inorganic materials.
\newblock \emph{npj Computational Materials}, 2\penalty0 (1):\penalty0 1--7,
  2016.

\bibitem[Wexler et~al.(2021)Wexler, Gautam, Stechel, and
  Carter]{wexler2021factors}
Robert~B Wexler, Gopalakrishnan~Sai Gautam, Ellen~B Stechel, and Emily~A
  Carter.
\newblock Factors governing oxygen vacancy formation in oxide perovskites.
\newblock \emph{Journal of the American Chemical Society}, 143\penalty0
  (33):\penalty0 13212--13227, 2021.

\bibitem[Woodward(1997)]{woodward1997octahedral}
Patrick~M Woodward.
\newblock Octahedral tilting in perovskites. i. geometrical considerations.
\newblock \emph{Acta Crystallographica Section B: Structural Science},
  53\penalty0 (1):\penalty0 32--43, 1997.

\bibitem[Xue and Liang(2017)]{xue2017robust}
Jingnan Xue and Faming Liang.
\newblock A robust model-free feature screening method for
  ultrahigh-dimensional data.
\newblock \emph{Journal of Computational and Graphical Statistics}, 26\penalty0
  (4):\penalty0 803--813, 2017.

\bibitem[Yamada et~al.(2008)Yamada, Takata, Hayashi, Shinohara, Azuma, Mori,
  Muranaka, Shimakawa, and Takano]{yamada2008perovskite}
Ikuya Yamada, Kazuhide Takata, Naoaki Hayashi, Satoshi Shinohara, Masaki Azuma,
  Shigeo Mori, Shigetoshi Muranaka, Yuichi Shimakawa, and Mikio Takano.
\newblock A perovskite containing quadrivalent iron as a
  charge-disproportionated ferrimagnet.
\newblock \emph{Angewandte Chemie}, 120\penalty0 (37):\penalty0 7140--7143,
  2008.

\bibitem[Zhao et~al.(2021)Zhao, Gao, Li, Qian, Shen, Wang, Shen, Hu, Dong,
  Huang, et~al.]{zhao2021combinatory}
Jianfa Zhao, Jiacheng Gao, Wenmin Li, Yuting Qian, Xudong Shen, Xiao Wang,
  Xi~Shen, Zhiwei Hu, Cheng Dong, Qingzhen Huang, et~al.
\newblock A combinatory ferroelectric compound bridging simple {ABO3} and
  a-site-ordered quadruple perovskite.
\newblock \emph{Nature communications}, 12\penalty0 (1):\penalty0 1--9, 2021.

\bibitem[Zhao et~al.(2012)Zhao, Liu, Roeder, Lafferty, and
  Wasserman]{zhao2012huge}
Tuo Zhao, Han Liu, Kathryn Roeder, John Lafferty, and Larry Wasserman.
\newblock The huge package for high-dimensional undirected graph estimation in
  r.
\newblock \emph{The Journal of Machine Learning Research}, 13\penalty0
  (1):\penalty0 1059--1062, 2012.

\bibitem[Zhou(2020)]{zhou2020structural}
J-S Zhou.
\newblock Structural distortions in rare-earth transition-metal oxide
  perovskites under high pressure.
\newblock \emph{Physical Review B}, 101\penalty0 (22):\penalty0 224104, 2020.

\end{thebibliography}


\end{document}